\documentclass[aps,twocolumn,preprintnumbers,amsmath,amssymb,nofootinbib,superscriptaddress,notitlepage]{revtex4-1}

\usepackage{epsfig}
\usepackage[utf8]{inputenc}

\usepackage{color}
 
\usepackage{hyperref}
\usepackage{multirow}

\begin{document}

\title{Heavy- and light-flavor symmetry partners of the $T_{cc}^+(3875)$,
  the $X(3872)$ \\ and the $X(3960)$ 
  from light-meson exchange saturation}

\author{Fang-Zheng Peng}
\affiliation{School of Physics,
Beihang University, Beijing 100191, China} 

\author{Mao-Jun Yan}
\affiliation{CAS Key Laboratory of Theoretical Physics, 
  Institute of Theoretical Physics, \\
  Chinese Academy of Sciences, Beijing 100190, China}

\author{Manuel Pavon Valderrama}\email{mpavon@buaa.edu.cn}
\affiliation{School of Physics, 
Beihang University, Beijing 100191, China} 

\date{\today}


\begin{abstract} 
  \rule{0ex}{3ex}
  The spectrum of the charmed meson-(anti)meson system is a fundamental tool
  for disentangling the nature of a few exotic hadrons, including
  the recently discovered $T_{cc}^+(3875)$ tetraquark, the $X(3960)$,
  or the $X(3872)$, the nature of which is still not clear after
  almost two decades of its discovery.
  Here we consider that the charmed meson-(anti)meson short-range interaction
  is described by the exchange of light-mesons
  ($\sigma$, $\rho$, $\omega$).
  The effects of light-meson exchanges are recast into a simple
  contact-range theory by means of a saturation procedure,
  resulting in a compact description of
  the two-hadron interaction.
  From this, if the $T_{cc}^+$ were to be an isoscalar $D^* D$ molecule,
  then there should exist an isoscalar $J=1$ $D^* D^*$ partner,
  as constrained by heavy-quark spin symmetry.
  Yet, within our model, the most attractive two charmed meson configurations
  are the isovector $J=0$ $D^* D^*$ molecule and its sextet $D_s^* D^*$
  and $D_s^* D_s^*$ flavor partners.
  Finally, we find a tension between the molecular descriptions of
  the $T_{cc}^+$ and that of the $X(3872)$ and $X(3960)$, where
  most parameter choices suggest that if the $T_{cc}^+$ is purely molecular
  then the $X(3872)$ overbinds (or conversely, if the $X(3872)$
  is a molecule the $T_{cc}^+$ does not bind).
  This might be consequential for determining the nature of these states.
\end{abstract}

\maketitle

\section{Introduction}

The LHCb has recently observed a state within the $D_s^+ D_s^{-}$ invariant
mass distribution of the $B^+ \to D_s^+ D_s^{-} K^+$ decay~\cite{LHCb:2022vsv}.
With a mass and width of
\begin{eqnarray}
  M &=& 3956 \pm 5 \pm 10 \,{\rm MeV} \, , \nonumber \\
  \Gamma &=& 43 \pm 13 \pm 8 \,{\rm MeV} \, ,
\end{eqnarray}
it has been named the $X(3960)$.
Its preferred quantum numbers are $J^{PC} = 0^{++}$ and it might be the same
resonance as the previous $X(3930)$ state observed in the $D^+ D^-$
invariant mass distribution~\cite{LHCb:2020pxc}, which in turn could
also happen to be the same state as the $\chi_0(3915)$ listed
in the Review of Particle Physics (RPP)~\cite{Workman:2022ynf}~\footnote{Or
  it might not: a recent study~\cite{Ji:2022vdj} suggests that
the $X(3930)$/$\chi_0(3915)$ is a $J^{PC} = 2^{++}$ state and
thus not identical to the $X(3960)$, while Ref.~\cite{Abreu:2023rye}
proposes a possible method to differentiate whether the $X(3930)$ and
$X(3960)$ are the same or not.}.
The $X(3960)$ can be interpreted as a charmed meson-antimeson
state~\cite{Ji:2022uie,Xin:2022bzt,Xie:2022lyw,Mutuk:2022ckn,Chen:2023eix},
from which it is natural to consider whether it could be related
to the well-known $X(3872)$, with $J^{PC} = 1^{++}$ and
whose mass and width are~\cite{Workman:2022ynf}
\begin{eqnarray}
  M &=& 3871.65 \pm 0.06 \, {\rm MeV} \, , \nonumber \\
  \Gamma &=& 1.19 \pm 0.21\,{\rm MeV} \, ,
\end{eqnarray}
which has been conjectured to be a $D^* \bar{D}$ bound state~\cite{Tornqvist:2003na,Voloshin:2003nt,Braaten:2003he}, though no consensus exists yet
on whether they are molecular or not~\cite{Swanson:2004pp,Dong:2009uf,Gamermann:2009fv,Gamermann:2009uq,Hanhart:2011tn,Esposito:2020ywk,Braaten:2020iqw,Guo:2022crh,Agaev:2022pis}.

If molecular,
the previous $X(3872)$ and $X(3960)$ states would be in principle related
with the doubly charmed tetraquark --- the $T_{cc}^+(3875)$ -- discovered 
in 2021 by the LHCb collaboration~\cite{LHCb:2021vvq,LHCb:2021auc} and
suspected to be a two charmed meson
bound state~\cite{Meng:2021jnw,Agaev:2021vur,Ling:2021bir,Dong:2021bvy,Feijoo:2021ppq,Fleming:2021wmk,Agaev:2022ast},
though predictions of the $T_{cc}^+$ as a compact tetraquark predate its
observation by decades~\cite{Carlson:1987hh,Silvestre-Brac:1993zem,Semay:1994ht,Pepin:1996id}.
The $T_{cc}^+$ is extremely close to the $D^{0*} D^{+}$
threshold, where the mass difference $\delta m = m(T_{cc}^+) - m(D) - m(D^*)$ is
\begin{eqnarray}
  \delta m_{\rm BW} = - 273 \pm 61 \pm 5 \, {}^{+11}_{-14} \,{\rm keV} \, ,
\end{eqnarray}
if a standard Breit-Wigner shape is used to fit
the $T_{cc}^+$~\cite{LHCb:2021vvq}, or
\begin{eqnarray}
  \delta m_{\rm U} = - 360 \pm 40 {}^{+4}_{-0} \,{\rm keV} \, ,
  \label{eq:delta-m-U}
\end{eqnarray}
if it is fitted with a unitarized Breit-Wigner shape~\cite{LHCb:2021auc},
where this later determination is probably more suitable for a state
close to a threshold.
This tetraquark is also relatively narrow, with a width of
\begin{eqnarray}
  \Gamma_{\rm BW} &=& 410 \pm 165 \pm 43 \, {}^{+18}_{-38} \,{\rm keV} \, , \\
  \Gamma_{\rm U} &=& 48 \pm 2 {}^{+0}_{-12} \,{\rm keV} \, , 
\end{eqnarray}
depending on the resonance profile used (where the unitarized BW is usually
considered the most reliable of the two).

Here we will consider the $X(3872)$, $X(3960)$ and $T_{cc}^+$
from the point of view of molecular spectroscopy and
within the particular phenomenological model
we proposed in~\cite{Peng:2021hkr}.
The questions we would like to address are:
Is it sensible to believe they are bound states?
Can the $T_{cc}^+$ be described with the same parameters
as the $X(3872)$ and $X(3960)$? What are their partner states?

For answering these questions
we will use a simple saturation model in which the effect of the
light-meson exchange forces is encapsulated in the coupling constant
of a contact-range theory, modulo a proportionality constant
that can be determined from a {\it reference state},
i.e. an observed state whose nature we will assume to be a specific type
of two-body bound state.
The idea of a reference state can in turn be used to relate different
molecular candidates.
This is particularly relevant to the connection between the $T_{cc}^+$ and
the $X(3872/3960)$: if these two states can be described with the same
set of parameters within a given model, this will give credence to
the idea that both of them conform to the assumptions
of said model.

However, as we will see, this is not the case in our model, where there is
a tension between a pure molecular explanation of the $T_{cc}^+$ as
a two charmed meson bound state and the $X(3872/3960)$
as charmed meson-antimeson systems.
That is, while the $X(3872)$ and $X(3960)$ can be easily explained
as $D^* \bar{D}$ and $D_s \bar{D}_s$ bound or virtual states
with the same parameters, this will lead to insufficient attraction
in the $D^* D$ system as to guarantee binding (though there will be
remarkable attraction nonetheless).
This result would be consistent with the picture of Janc and
Rosina~\cite{Janc:2004qn}, who conjectured that the binding of a tetraquark
below the $D^* D$ threshold requires both the molecular and quark components.

\section{Saturation of the contact-range couplings}
\label{sec:saturation}

We will consider a generic two-hadron system $H_1 H_2$, which we will describe
by means of an S-wave contact-range theory containing a central
and spin-spin term
\begin{eqnarray}
  V_C = C = C_0 + C_1 \,\hat{\vec{S}}_{L1} \cdot \hat{\vec{S}}_{L2} \, ,
\end{eqnarray}
where $V_C$ is the contact-range potential in p-space, $C$ is its strength
(or coupling), which can be subdivided into $C_0$ and $C_1$,
the central and spin-spin coupling constants, and
$\hat{\vec{S}}_{Li} = \vec{S}_{Li} / | S_{Li} |$ is the {\it reduced}
light-spin operator for hadron $i=1,2$ (where $\vec{S}_{Li}$ is the
light-spin operator and $S_{Li}$ the total light-spin of
hadron $i=1,2$).
For S-wave charmed mesons, $\hat{\vec{S}}_{Li} = \vec{\sigma}_{Li}$, with
$\vec{\sigma}$ representing the Pauli-matrices.

Here it is interesting to notice that the dependence on the light-spin
(i.e. the spin of the light $q = u,d,s$ quarks within hadrons
$H_1$ and $H_2$) is a consequence of heavy-quark spin
symmetry (HQSS) which implies that any dependence on the spin of heavy quarks
will be suppressed by a factor of $\Lambda_{\rm QCD} / m_Q$,
with $\Lambda_{\rm QCD} \approx 200\,{\rm MeV}$.
In principle there could be higher spin operators (quadrupolar, octupolar,
etc.), but in practice higher order terms will be increasingly suppressed.
Besides, for S-wave charmed mesons we have $S_{L} = {1}/{2}$, which precludes
the appearance of the aforementioned high spin operators.

The contact-range approximation assumes that the binding momentum of
the two hadrons is not high enough as to disentangle
the details of the interaction binding them.
That is, we do not have to consider the full two-hadron potential and
all its details, but only its coarse-grained properties.
We will do this by assuming that the two-hadron potential is derived from
the exchange of light-mesons ($\sigma$, $\rho$, $\omega$), which will then
saturate the coupling constants.
Finally, pions are long-ranged and consequently do not saturate
the contact-range couplings of our model.
They have also been shown to be perturbative for systems composed of
two charmed mesons~\cite{Valderrama:2012jv,Nieves:2012tt}, a point
that we have also explicitly checked with concrete calculations
in our saturation model~\cite{Peng:2021hkr}.
Thus we will simply ignore their contribution, as it can be safely neglected.

For the saturation of the couplings, we will consider the exchange of
the scalar and vector mesons separately, as these mesons
have different masses.
For the scalar meson (the $\sigma$), the potential takes the form
\begin{eqnarray}
  V_S(\vec{q}\,) = - \frac{g_{S1} g_{S2}}{m_S^2 + {\vec{q}\,}^2} \, ,
\end{eqnarray}
where $\vec{q}\,$ is the exchanged momentum, $m_S$ is the scalar meson mass and
$g_{Si}$ with $i=1,2$ is the scalar coupling for hadrons $H_1$ and $H_2$.
This leads to the following contribution to the couplings
\begin{eqnarray}
  C_0^S(\Lambda \sim m_S) &\propto& -\frac{g_{S1} g_{S2}}{m_S^2} \, , \\
  C_1^S(\Lambda \sim m_S) &\propto& 0 \, ,
\end{eqnarray}
where we remind that saturation works best for a regularization scale of
the same order of magnitude as the mass of the exchanged particle (hence
$\Lambda \sim m_S$) and that we only expect the contact-range couplings
to be proportional to the potential at zero momentum, modulo an unknown
proportionality constant.
For the vector meson, the potential is
\begin{eqnarray}
  V_V(\vec{q}\,) &=& \frac{g_{V1} g_{V2}}{m_V^2 + {\vec{q}\,}^2}
  + \frac{f_{V1} f_{V2}}{6 M^2}\,\frac{m_V^2}{m_V^2 + {\vec{q}\,}^2}\,
  \hat{\vec{S}}_{L1} \cdot \hat{\vec{S}}_{L2}  \, \nonumber \\
  &+& \dots \, ,
  \label{eq:V-V}
\end{eqnarray}
where we have momentarily ignored isospin, SU(3)-flavor and G-parity factors
for obtaining a more compact expression; $m_V$ is the vector meson mass,
$g_{Vi}$ the electric-type (E0) coupling, $f_{Vi}$ the magnetic-type (M1)
coupling, and $M$ a scaling mass for the magnetic-type term for which
we will choose the nucleon mass $M = m_N = 938.9\,{\rm MeV}$.
The dots represent S-to-D wave terms and Dirac-delta contributions that
do not saturate of our contact-range potential~\cite{Peng:2020xrf}.
The previous potential leads to the contributions
\begin{eqnarray}
  C_0^V(\Lambda \sim m_V) &\propto&
  \frac{g_{V1} g_{V2}}{m_V^2}\,(\zeta + T_{12}) \, , \label{eq:C0-V} \\
  C_1^V(\Lambda \sim m_V) &\propto&
  \frac{f_{V1} f_{V2}}{6 M^2}\,(\zeta + T_{12}) \, , \label{eq:C1-V} 
\end{eqnarray}
where we have now included the isospin and G-parity factors for two charmed
meson systems without strangeness, with $\zeta = +1$ ($-1)$
for the meson-meson (meson-antimeson) case and
$T_{12} = \vec{\tau}_1 \cdot \vec{\tau}_2$
the isospin factor.
The extension to systems with strangeness is straightforward, as it only
requires to consider the appropriate SU(3)-flavor Clebsch-Gordan coefficients.

Next, we have to combine the contributions from scalar and vector meson
exchanges, where there is the problem that the respective renormalization
scales at which saturation is optimal is different
for the scalar and vector mesons.
For this we will use a renormalization group equation (RGE) to determine
how the couplings evolve when the regularization scale $\Lambda$
changes.
For non-relativistic two-body theories, the RGE of a contact-range coupling
$C(\Lambda)$ takes the form~\cite{PavonValderrama:2014zeq}
\begin{eqnarray}
  \frac{d}{d \Lambda} \langle \Psi | V_C(\Lambda) | \Psi \rangle = 0 \, ,
\end{eqnarray}
with $V_C(\Lambda)$ the regularized contact-range potential,
$\Lambda$ the regularization scale and
$\Psi$ the two-body wavefunction.
If the wavefunction displays a power-law behavior of the type
$\Psi(r) \propto r^{\alpha/2}$ at distances $r \sim 1/\Lambda$,
the RGE will simplify to~\cite{PavonValderrama:2014zeq}
\begin{eqnarray}
  \frac{d}{d \Lambda}\,
  \left[  \frac{C(\Lambda)}{\Lambda^{\alpha}} \right] = 0 \, ,
\end{eqnarray}
or, equivalently
\begin{eqnarray}
  \frac{C(\Lambda_1)}{\Lambda_1^{\alpha}} = \frac{C(\Lambda_2)}{\Lambda_2^{\alpha}} \, ,
  \label{eq:RG-comparison}
\end{eqnarray}
with $\Lambda_1$ and $\Lambda_2$ two values of the regularization scale.
For the particular case at hand, i.e. the saturated couplings at the scalar and
vector mass scales, the previous RGE leads to the following combination
\begin{eqnarray}
  C(m_V) = C_V(m_V) + {\left(\frac{m_V}{m_S}\right)}^{\alpha} \, C_S(m_S) \, .
  \label{eq:combination}
\end{eqnarray}
For determining the exponent $\alpha$ we are still required to make
assumptions about the power-law properties of the wave function
at distances $m_V r \sim 1$, i.e., $\Psi(r) \sim r^{\alpha/2}$.
From the semi-classical approximation and the Langer correction~\cite{Langer:1937qr}, we end up with $\alpha = 1$.

Putting all the pieces together, the saturated contact-range coupling
is determined modulo an unknown proportionality constant and
takes the following form
\begin{eqnarray}
  && C^{\rm sat}(\Lambda = m_V) =
  C_0^{\rm sat} + C_1^{\rm sat}\,\hat{\vec{S}}_{L1} \cdot \hat{\vec{S}}_{L2}
  \nonumber \\
  && \quad \propto 
  \frac{g_{V1} g_{V2}}{m_V^2}
  \left[ \zeta + \hat{T}_{12} \right]
  \left( 1 + \kappa_{V1} \kappa_{V2} \, \frac{m_V^2}{6 M^2}\,\hat{C}_{L12} \right)
  \nonumber \\
  && \quad \,\, - {(\frac{m_V}{m_S})}^{\alpha}\frac{g_{S1} g_{S2}}{m_S^2} \, , 
    \label{eq:coupling-sat}
\end{eqnarray}
where $\hat{T}_{12} = \hat{\vec{I}}_1 \cdot \hat{\vec{I}}_2$,
$\hat{C}_{L12} = \hat{\vec{S}}_{L1} \cdot \hat{\vec{S}}_{L2}$
and $\kappa_{Vi} = f_{Vi} / g_{Vi}$.

\section{Calibration}
\label{sec:calibration}

In this section we calibrate the RG-saturation model.
This requires to specify the regularization of the contact-range potential and
the masses and couplings of the light mesons.
Yet, the most important factor to make concrete predictions is
the determination of the proportionality constant for the saturated couplings.
For this we have to choose a reference state --- a plausible molecular
candidate --- from which to derive all other predictions.
We will consider a few candidates --- the $X(3872)$, $X(3960)$,
$T_{cc}^+(3875)$ and the $D\bar{D}$ state calculated
in the lattice~\cite{Prelovsek:2020eiw} --- and argue
that the $X(3872)$ as a $1^{++}$ $D^*\bar{D}$ bound state is
the most suitable choice from which to derive the molecular spectrum of
the two charmed meson systems.

\subsection{Calculating the spectrum}

For the regularization, we use a regulator function $f(x)$
\begin{eqnarray}
  V_C = C^{\rm sat} f(\frac{p'}{\Lambda}) f(\frac{p}{\Lambda}) \, ,
  \label{eq:Vc-reg}
\end{eqnarray}
for which we will choose a Gaussian, $f(x) = e^{-x^2}$, and where $\Lambda$
is the cutoff, which we take to be $\Lambda = 1\,{\rm GeV}$
as in~\cite{Peng:2021hkr}.
This potential will be plugged into the Lippmann-Schwinger equation
for a bound state
\begin{eqnarray}
  \phi(p) + 2\mu\,\int \frac{d^3 \vec{q}}{(2 \pi)^3}\,
  \frac{\langle p | V_C | q \rangle}{M_{\rm p} - \frac{q^2}{2 \mu} - M_{\rm th}}
  \,\phi(q) = 0 \, ,
\end{eqnarray}
where $\phi(p)$ refers to the vertex function, which is related to
the wavefunction $\Psi(p)$ by the relation
$(M_{\rm p} - \frac{p^2}{2 \mu} - M_{\rm th}) \Psi(p) = \phi(p)$ ,
$\mu$ is the reduced mass of the system,
while $M_{\rm th}$, $M_{\rm p}$ are the mass of the threshold and of
the pole of the two-body scattering amplitude (i.e. the mass of
the bound state), respectively.
In the case at hand, this equation further simplifies to
\begin{eqnarray}
  1 + \frac{\mu}{4 \pi^2} \,C^{\rm sat}(\Lambda) I(\gamma_2, \Lambda) = 0 \, ,
\end{eqnarray}
where $\gamma_2 = \sqrt{2 \mu (M_{\rm th }- M_{\rm p})}$ is the wave number of
the bound state and with $I(\gamma_2, \Lambda)$ given by
\begin{equation}
  I(\gamma_2, \Lambda) = \sqrt{2 \pi}\,\Lambda -
  2\, e^{2 \gamma_2^2 / \Lambda^2}\,\pi \gamma_2 \,
  {\rm erfc}\left( \frac{\sqrt{2} \gamma_2}{\Lambda} \right) \, ,
\end{equation}
where ${\rm erfc}\,(x)$ is the complementary error function.

For calibrating the $C^{\rm sat}(\Lambda)$ coupling and its unknown
proportionality constant in Eq.~(\ref{eq:coupling-sat})
we will use a reference state.
In the doubly charmed sector the only candidate state is the $T_{cc}^+$
tetraquark, while in the hidden charmed sector besides the experimentally
observed $X(3872)$ we may also include the recent $X(3960)$ (for which
we will use the location of its pole in the scattering amplitude
as determined for instance in~\cite{Ji:2022uie}) or the $D\bar{D}$ and
$D_s \bar{D}_s$ poles found in the lattice~\cite{Prelovsek:2020eiw}.
As the binding energy of the reference state is known, we just
obtain $C^{\rm sat}_{\rm ref}$ from the bound state equation
\begin{eqnarray}
  1 + \frac{\mu_{\rm ref}}{4 \pi^2}\,C^{\rm sat}_{\rm ref}\,
  I(\gamma_2, \Lambda) = 0 \, . \label{eq:calibration}
\end{eqnarray}
After this, if we want to predict the mass of a given molecule,
we calculate the ratio
\begin{eqnarray}
  R_{\rm mol} =
  \frac{\mu_{\rm mol} C^{\rm sat}_{\rm mol}}{\mu_{\rm ref} C^{\rm sat}_{\rm ref}} \, ,
  \label{eq:Rmol}
\end{eqnarray}
which is independent of the proportionality constant, and then from $R_{\rm mol}$
we simply solve the bound state equation
\begin{eqnarray}
  1 + R_{\rm mol}\,(\frac{\mu_{\rm ref}}{4 \pi^2}\,C^{\rm sat}_{\rm ref})\,
  I(\gamma_2, \Lambda) = 0 \, ,
\end{eqnarray}
to obtain the binding energies.

For the couplings of the light-meson with the charmed mesons, we will resort
to a series of well-known phenomenological relations.
For the vector mesons ($\rho$, $\omega$, $K^*$ and $\phi$)
we have simply made use of the mixing of these mesons with the electromagnetic
current (vector meson
dominance~\cite{Sakurai:1960ju,Kawarabayashi:1966kd,Riazuddin:1966sw})
as a way to determine the $g_V$ and $\kappa_V$ (E0 and M1)
couplings: we can match $g_V$ and $\kappa_V$ to the charge and
magnetic moment of the particular hadron we are interested in.
This results in $g_V = 2.9$ and $\kappa_V = 2.85$,
as explained in~\cite{Peng:2021hkr}.
For the scalar meson the linear sigma model~\cite{GellMann:1960np}
predicts $g_{SNN} = \sqrt{2} m_N / f_{\pi} \simeq 10.2$ for the nucleon,
where $m_N$ is the nucleon mass and $f_{\pi} \simeq 132\,{\rm MeV}$
the pion weak decay constant.
For the charmed meson, which contains one light-quark instead of three,
we assume the quark model relation $g_S = g_{S qq} = g_{S NN} / 3 = 3.4$, i.e.
that the coupling of the sigma is proportional to the number of
light-quarks within the hadron.

In the strange sector the contributions coming from the exchange of
the $K^*$ and $\phi$ vector mesons have a shorter range than
the ones from the $\rho$ and $\omega$ mesons.
In analogy to Eq.~(\ref{eq:combination}), their contribution will be slightly
suppressed in comparison to the non-strange vector mesons.
If we take the $D_s^{(*)} D_s^{(*)}$ and $D_s^{(*)} \bar{D}_s^{(*)}$ systems
as an illustrative example, the saturated coupling will be 
\begin{eqnarray}
  C^{\rm sat}_{\rm mol} &\propto&
  {(\frac{m_V}{m_\phi})}^{\alpha}
    \frac{g_{V}^2}{m_{\phi}^2}
  \left[ 2 \zeta \right]
  \left( 1 + \kappa_{V}^2 \,
  \frac{m_{\phi}^2}{6 M^2}\,\hat{C}_{L12} \right)
  \nonumber \\
  && \quad \,\, - {(\frac{m_V}{m_S})}^{\alpha}\frac{g_{S}^2}{m_S^2} \, , 
\end{eqnarray}
where we follow the conventions of Eq.~(\ref{eq:combination}),
except for the couplings for which we use the values
mentioned in the previous paragraph.
As can be appreciated, the $\phi$ vector meson contribution is suppressed by
a $(m_V/m_{\phi})^{\alpha}$ factor with respect to the $\rho$ and
$\omega$ contributions that we use as our baseline.
For the masses of the vector mesons we use
$m_V = (m_\rho + m_\omega)/2 = 775\,{\rm MeV}$,
$m_{K^*} = 890\,{\rm MeV}$ and $m_{\phi} = 1020\,{\rm MeV}$.
The mass of the scalar meson is taken to be $m_S = 475\,{\rm MeV}$,
i.e. the middle value of the $(400-550)\,{\rm MeV}$ range listed
in the RPP~\cite{Workman:2022ynf}.

We also assume that the coupling of the scalar
to the $s$-quark is approximately the same as to the $u$- and
$d$-quarks: $g_S = g_{S uu} = g_{S dd} = g_{S ss}$.
This assumption works well when comparing the $D\bar{D}$ and $D_s \bar{D}_s$
systems predicted in the lattice~\cite{Prelovsek:2020eiw}:
a good prediction of the $D_s \bar{D}_s$ from the $D \bar{D}$ (or vice versa)
requires the coupling of the sigma to be similar to the strange
and non-strange quarks.
Indeed if we use the location of the $D\bar{D}$ bound state as input,
we predict the $D_s \bar{D}_s$ state to be located at
\begin{eqnarray}
  M_p - 2 m(D_s) = {(-1.0)}^V \, [0.2 - 0.5 \,i] \, {\rm MeV} \, ,
\end{eqnarray}
where the first (parentheses) and second (brackets) value
correspond to a single ($D_s \bar{D}_s$) and coupled
($D\bar{D}$-$D_s\bar{D}_s$-$D^*\bar{D}^*$) channel calculation.
Notice that while in the single channel case we obtain a virtual state close
to the $D_s\bar{D}_s$ threshold (which we indicate with the $^V$ superscript),
in the coupled channel calculation the pole is located
in the (I,II) Riemann sheet of channels $D\bar{D}$ and $D_s\bar{D}_s$,
respectively.
This is to be compared with $M_p - 2 m(D_s) = - 0.2^{+0.16}_{-4.9} - 0.27^{+2.5}_{-0.15} \frac{i}{2}$ in~\cite{Prelovsek:2020eiw}, which is located in sheets (II,I)
for $70\%$ of the bootstrap samples or in sheets (I,II) for the rest.
We note that this choice is also necessary for reproducing the $Z_{cs}(3985)$
as a $D^* \bar{D}_s$-$D \bar{D}_s^*$ molecule, as we previously discussed
in Ref.~\cite{Yan:2021tcp}.

It is also interesting to notice that the calculation above is not only
relevant for the choice of the scalar coupling with the s-quark,
but also for the relative importance of coupled channel effects.
Indeed, the difference in the position of the $D_s \bar{D}_s$ pole is merely of
the order of $1\,{\rm MeV}$ for the single and coupled channel cases,
which suggest a minor role for the coupled channel dynamics.
This conclusion agrees with our previous numerical results
in~\cite{Peng:2021hkr}.
Besides, the $\mathcal{O}(1\,{\rm MeV})$ estimation for the coupled channel
effects happen to be smaller than the uncertainties of our model,
which we explain in the next few lines.
For the previous reasons we will not further consider coupled channel effects
when calculating the molecular spectrum.

\subsection{Error estimations}
\label{subsec:errors}

For calculating the expected errors of the model we will consider two sources of
uncertainty.
The first is the light-meson exchange potential itself,
for which the parameters are not that well-known.
In principle this would include a lot of uncertainties
coming from each individual parameter, but in practice the single largest
source of uncertainty is the scalar meson.
Its large width and the uncertainty in its mass are both well-known issues
within the light meson exchange picture, for which several solutions
exists~\cite{Machleidt:1987hj,Machleidt:1989tm,Binstock:1971duy,Flambaum:2007xj}
(for a more detailed explanation,
we refer to Appendix B of~\cite{Peng:2021hkr}).
Here in particular we consider the finding that the exchange of a wide meson
can be effectively approximated by a narrow one after a redefinition of
its parameters~\cite{Machleidt:1987hj,Machleidt:1989tm}.

For modeling the error derived from the scalar meson
we will simply vary its mass within the range listed
in the RPP~\cite{Workman:2022ynf}, i.e. $400-550\,{\rm MeV}$, leading to
\begin{eqnarray}
  \Delta M_{\rm OBE} = M_p(m_S \pm \Delta m_S) - M_p(m_S) \, ,
\end{eqnarray}
with $m_S = 475 \pm 75\,{\rm MeV}$, where this error is asymmetric.
Numerically, this is equivalent to changing the magnitude of $C^{\rm sat}$ by
up to $30-40\%$ depending on the specific molecular configuration,
which is for instance comparable with the relative errors used
for the light-meson exchange potential within
the OBE model of~\cite{Liu:2019stu}.

The second source of uncertainty is the contact-range approximation itself,
the accuracy of which depends on the comparison between the momentum
scale of the predicted bound state and the masses of the exchanged
mesons. That is, we will assume a relative error of
size $\gamma_2 / m_V$:
\begin{eqnarray}
  \Delta M_{\rm contact} = B_{\rm mol} \, \left( \frac{\gamma_2}{m_V} \right) \, ,
\end{eqnarray}
which happens to be symmetric (and in most cases smaller than the error
coming from the uncertainty in the scalar meson mass) and
where $B_{\rm mol} = M_{\rm th} - M_{p}$ refers to the binding energy.
Finally, we will sum these two errors in quadrature.

\subsection{Reference states}

For the reference state we have considered four possible choices:
the $X(3872)$, the $X(3960)$, the $T_{cc}^+(3875)$ and
the $0^{++}$ $D\bar{D}$ bound state
found in the lattice~\cite{Prelovsek:2020eiw}.
The specific inputs we will use are:
\begin{itemize}
\item[(a)] For the $X(3872)$ we will consider it as a $1^{++}$ $D^*\bar{D}$ bound state
with the mass as determined in the RPP~\cite{Workman:2022ynf},
i.e. $3871.65 \pm 0.06\,{\rm MeV}$ which we round up to
$3871.7\,{\rm MeV}$.

\item[(b)] For the $X(3960)$ we will refer to the recent theoretical analysis
of~\cite{Ji:2022uie}, which considers the $X(3960)$ to be either
a virtual or bound $0^{++}$ $D_s\bar{D}_s$ state.
The determination of the pole mass in Ref.~\cite{Ji:2022uie} happens to be
the same in the virtual and bound states cases,
yielding $3928 \pm 3 \,{\rm MeV}$.
However, we will only consider the virtual state solution as the input
for our calculations. This is because other input choices already
predict a bound $X(3960)$ in the same mass range of Ref.~\cite{Ji:2022uie}.

\item[(c)] For the $T_{cc}^+(3875)$ we will use its mass as determined from
  the unitarized Breit-Wigner shape in~\cite{LHCb:2021auc}, i.e. by
  the $\delta m$ in Eq.~(\ref{eq:delta-m-U}).
  As we are using the isospin symmetric limit, this will give a bound state
  energy of $B = 1.065\,{\rm MeV}$.

\item[(d)] For the $0^{++}$ $D\bar{D}$ bound state we use the binding energy
  calculated in the lattice~\cite{Prelovsek:2020eiw},
  i.e. $B = 4.0^{+5.0}_{-3.7}\,{\rm MeV}$.
\end{itemize}
We will not propagate the error in the input masses, as they usually generate
smaller uncertainties than the scalar meson or the contact-range approximation.
For choosing which one of these four inputs to use,
we will compare the predictions derived from each of the inputs
for the spectroscopy of a few molecular candidates and
for the decays of the $T_{cc}^+$ tetraquark,
as we will explain in the following lines.

\subsection{Spectroscopy comparison}

First, we consider the predictions for the spectroscopy of the charmed
meson-(anti)meson molecular candidates depending on whether
the reference state is the $X(3872)$, the $X(3960)$, the $T_{cc}$ or
the $D\bar{D}$ state found in the lattice.
We list these predictions in Table~\ref{tab:basic-predictions}
for the molecular configurations for which there exists a clear candidate
state (observed either in the experiment or the lattice).

The reason for comparing the predictions for different inputs is that
we want to check the internal consistency of the molecular hypothesis
for the most common candidates in our model.
What we find is that the $X(3872)$ and $X(3960)$ can be explained with the same
set of parameters within the uncertainties of our model,
but this is not the case for the $T_{cc}$.
This indicates that if the $X(3872)$ or the $X(3960)$ were to be explained
in purely molecular terms, the $T_{cc}$ would require additional
attraction to bind, which might very well come from non-molecular components.
Alternatively, were the $T_{cc}$ to be purely molecular, then it would be
difficult to explain the mass of the $X(3872)$ and $X(3960)$
within the molecular picture in our saturation model.

We elaborate this argument in more detail in the following lines:
\begin{itemize}
\item [(i)] If we use the $X(3872)$ as the reference state, there are
  a few predictions in Table~\ref{tab:basic-predictions}
  worth commenting:
  \begin{itemize}
  \item[(i.a)] There is a $0^{++}$
  $D_s \bar{D}_s$ virtual state at $3922^{+11}_{-15}\,{\rm MeV}$,
  which might be identified with the $X(3960)$.
  Even though the $X(3960)$ has been found at $3956\,{\rm MeV}$,
  i.e. above the $D_s \bar{D}_s$ threshold, this determination of
  its mass depends on the assumption that the $X(3960)$ is correctly
  described with a Breit-Wigner profile. As has been 
  discussed in the literature~\cite{Hanhart:2015cua,Dong:2020hxe},
  this is not necessarily true for bound and virtual states close to threshold.
  In this regard, Ref.~\cite{Ji:2022uie} recently analyzed the $D_s^+ D_s^{-}$
  invariant mass distribution of~\cite{lhcb2022-a} and found that
  the $X(3960)$ can be described either as a virtual or
  bound state at about $3928 \pm 3\,{\rm MeV}$, which is compatible
  with our prediction in the virtual state case (but not if
  the $X(3960)$ turns out to be a bound state).

  \item[(i.b)] However, if we turn our attention to the $T_{cc}^+$,
  there is not enough attraction to bind the $D^*D$ system
  (though it is still attractive).
  In this case, the $T_{cc}^+$ will have to receive additional attraction
  from its short-range quark degrees of freedom
  in order to be able to bind.
  This scenario would be compatible with the seminal calculation by Janc and
  Rosina~\cite{Janc:2004qn}, which considered both the quark and meson
  components of the $T_{cc}^+$ as necessary for generating this state.
  There is a series of predictions~\cite{Silvestre-Brac:1993zem,Semay:1994ht,Maiani:2019lpu,Luo:2017eub} that put the mass of
  the $T_{cc}^+$ tetraquark above the $D^* D$ threshold,
  in which case this compact component should be able to mix
  with the molecular component and provide additional
  attraction not taken into account in our model.

\item[(i.c)] Besides this, the mass of the $2^{++}$ partner of the $X(3872)$
  is $4011.6 \pm 0.8$, which is compatible with the mass and
  quantum numbers of a $2^{++}$
  resonance recently observed by Belle~\cite{Belle:2021nuv}
  ($M = 4014.3 \pm 4.0 \pm 1.5 \,{\rm MeV}$ and $J=0$ or $2$,
  though the signal has poor statistical significance).
  This $2^{++}$ partner, which was conjectured a decade
  ago~\cite{Valderrama:2012jv,Nieves:2012tt}, is a consequence
  of HQSS: if the $X(3872)$ is indeed a $1^{++}$ $D^*\bar{D}$ bound state,
  it should have a $2^{++}$ $D^*\bar{D}^*$ partner state.
  The location of this state is usually predicted at $4012-4013\,{\rm MeV}$
  when pion interactions are neglected and $4015\,{\rm MeV}$ if the effects
  of the one pion exchange (OPE) potential are included~\cite{Nieves:2012tt}.
  The width of this state has been previously estimated to be of the order of
  a few MeV~\cite{Albaladejo:2015dsa}
  (from $0.9$ to $14.0\,{\rm MeV}$ depending on the assumptions),
  which is in line with the width measured by Belle~\cite{Belle:2021nuv}
  ($\Gamma = 4 \pm 11 \pm 6\,{\rm MeV}$).
  
\item[(i.d)] We also obtain a $D\bar{D}$ virtual state close to threshold,
  which might correspond to the $D\bar{D}$ bound state found in the lattice,
  and we predict the $Z_c$ and $Z_{cs}$ states as $D^*\bar{D}$ and $D_s^*\bar{D}$
  virtual states, where the masses are in line with the previous
  theoretical analyses of Refs.~\cite{Albaladejo:2015lob,Yang:2020nrt}.
  \end{itemize}

\item[(ii)] If we use the $X(3960)$ as the reference state, and assume it to be
  a virtual state with the mass extracted in Ref.~\cite{Ji:2022uie},
  the predictions are basically compatible with the ones derived
  from the $X(3872)$ within errors (and thus we do not comment
  the results in detail)
  That is, provided the $X(3960)$ is a virtual state, its molecular
  interpretation will be compatible with that of the $X(3872)$.
  
\item[(iii)] If we use the $T_{cc}^+$ as the reference state ($D^*D$ molecule
  with $I = 0$), we find that:
  \begin{itemize}
  \item[(iii.a)] the $X(3872)$ is predicted a few dozen MeV below threshold.
  In this scenario the $X(3872)$ cannot be a pure molecular state within
  the uncertainties of our model, and there should be shorter range
  components (e.g. coupling to charmonium) that push
  the $D^* \bar{D}$ pole closer to threshold.

\item[(iii.b)] the $X(3960)$ is predicted a few MeV below threshold,
  in particular at $M = 3931.6^{+1.8}_{-2.7}\,{\rm MeV}$, which compares
  well with the bound state pole determination in~\cite{Ji:2022uie},
  $M = 3928 \pm 3\,{\rm MeV}$.
  \end{itemize}
\end{itemize}
From the previous inputs it is already evident that the molecular explanations
of the $X(3872)$ and $T_{cc}$ are not compatible within our model.
The situation is more ambiguous with respect to the $X(3960)$ though,
for which the virtual and bound state interpretations are compatible
with a molecular $X(3872)$ and $T_{cc}$, respectively.
The same comment applies to the $Z_c(3900)$ and $Z_{cs}(3985)$.
However, if the $2^{++}$ state observed by Belle is eventually confirmed and
happens to be the conjectured $X(4012)$, then this will favor (disfavor)
the interpretation of the $X(3872)$ ($T_{cc}$) as molecular.

If we also include the lattice results in our discussion, we find that
\begin{itemize}
\item[(iv)] Using the $X(3872)$ as input will lead to virtual states
  in the $D\bar{D}$ and $D_s\bar{D}_s$ systems that are close to threshold.
  They are not as attractive as in the lattice calculations, but still
  compatible with them within errors.
\item[(v)] Using the $T_{cc}$ as input will lead to a $D\bar{D}$ that is
  markedly more bound than in the lattice, though uncertainties are
  too large to claim that they are incompatible.
\item[(vi)] If we use the lattice results as the input, it happens that
  the location of the $D\bar{D}$ and $D_s\bar{D}_s$ poles are compatible
  with each other, but predict the $X(3872)$ to be too bound
  by tens of MeV.
  The $T_{cc}^+$ will not bind either in this scenario, but will still
  survive as a virtual state close to threshold. In this case,
  the importance of the quark degrees of freedom in binding
  will be smaller than when the $X(3872)$ is the reference
  state.
\end{itemize}
That is, this additional comparison does not provide clear-cut answers,
though it shows a weak preference for the $X(3872)$ (instead of
the $T_{cc}^+$) to be more molecular.

Regarding the conflict between the $X(3872)$ and $T_{cc}(3875)$, it is important
to notice the mixing between the meson-antimeson components of the $X(3872)$ and
the possible nearby $\chi_{c1}$ charmonium and the molecular $T_{cc}(3875)$
and its compact quark components.
This mixing can be parametrized in terms of a potential contribution
of the type
\begin{eqnarray}
  V =
  \begin{pmatrix}
    V_{\rm mol} & C_{\rm comp} \\
    C_{\rm comp} & 0
  \end{pmatrix} \, ,
\end{eqnarray}
which results in a mass shift proportional to
\begin{eqnarray}
  \Delta M_{\rm mol} \propto - \frac{C_{\rm comp}^2}{\Delta} \, ,
\end{eqnarray}
with $\Delta$ the mass difference between the (uncoupled) compact
and molecular components, $\Delta = M_{\rm comp} - M_{\rm mol}$.
This will make the molecular candidate lighter or heavier depending on
the location of the non-molecular component (before mixing).
That is, if the original predictions of the compact component are
heavier than the two-hadron threshold, the two-hadron system will increase
its attraction with respect to the purely molecular scenario.

It happens that most predictions of the $\chi_{c1}(2P)$ in pure charmonium
models are heavier~\cite{Godfrey:1985xj,Zeng:1994vj,Ebert:2002pp}
(usually in the $3.9-4.0\,{\rm GeV}$ ballpark)
than the $X(3872)$, from which the contribution of the compact
components is more likely to be attractive
than repulsive. 
The situation is more open for a compact $cc \bar{u} \bar{d}$ tetraquark, as
there is a comparable number of predictions below~\cite{Zouzou:1986qh,Vijande:2003ki,Lee:2009rt,Feng:2013kea},
above~\cite{Silvestre-Brac:1993zem,Semay:1994ht,Maiani:2019lpu,Luo:2017eub} or compatible with~\cite{Carlson:1987hh,Gelman:2002wf,Navarra:2007yw,Karliner:2017qjm,Maiani:2022qze} the $D^* D$ threshold.
This slightly favors scenarios in which the attraction in the $D^* D$ system
is by itself not enough as to generate a bound state below threshold
(as it happens to be probable that a pure $cc \bar{u} \bar{d}$ state
will provide the missing attraction),
while disfavors the scenarios in which the $D^* \bar{D}$ system
shows overbinding (as it is improbable that the closest pure $c\bar{c}$
state will happen to be lighter than the $X(3872)$).

\begin{table*}[!ttt]
\begin{tabular}{|c|cccccccc|}
\hline\hline
Input & System & $I$($J^{P(C)}$) & $S$ & $R_{\rm mol}$ & $B_{\rm mol}$ & $M_{\rm mol}$
& Candidate & $M_{\rm candidate}$\\
\hline \hline
       {\multirow{6}{*}{$X(3872)$}} &
$D^* {D}$ & $0$ ($1^+$) & $0$ & $0.45$ & ${(24.0)}^V$ & $3852^{+17}_{-24}$ & $T_{cc}$ & $3875.7$ \\
       \cline{2-9} &
$D \bar{D}$ & $0$ ($0^{++}$) & $0$ & $0.70$ & ${(1.5)}^V$ & $3733.0^{+1.2}_{-1.9}$ & - &
       $3730.5^{+3.7}_{-5.0}$~\cite{Prelovsek:2020eiw} \\
& $D_s \bar{D}_s$ & $0$ ($0^{++}$) & $0$ & $0.51$ & ${(15)^V}$ & $3922^{+11}_{-15}$ & $X(3960)$ &
       $3930^{+3.8}_{-2.0}$~\cite{Prelovsek:2020eiw},
       ${(3928 \pm 3)}^{V/B}$~\cite{Ji:2022uie} \\
       & $D^* \bar{D}$ & $0$ ($1^{++}$) & $0$ & $1.00$ & $4.1$ & Input & $X(3872)$ & $3871.7$ \\
& $D^* \bar{D}^*$ & $0$ ($2^{++}$) & $0$ & $1.04$ & $5.5$ & $4011.6 \pm 0.7$ & - & $4014.3 \pm 4.0 \pm 1.5$~\cite{Belle:2021nuv} \\
       \cline{2-9} &
$D^* \bar{D}$ & $1$ ($1^{+}$) & $0$ & $0.44$ & ${(26)}^V$ & $3850^{+19}_{-26}$ & $Z_c(3900)$ & ${(3831-3844)}^V$~\cite{Albaladejo:2015lob} \\
& $D^* \bar{D}_s$-$D \bar{D}_s^*$ & $\frac{1}{2}$ ($1^{+}$) & $-1$ & $0.45$ & ${(24)}^V$ & $3955^{+17}_{-25}$ & $Z_{cs}(3985)$ & ${(3971-3974)^V}$~\cite{Yang:2020nrt} \\
\hline\hline
    {\multirow{6}{*}{$X(3960)$}} &
    $D^* {D}$ & $0$ ($1^+$) & $0$ & $0.88$ & ${(16.2)}^V$ & $3859.6^{+4.5}_{-5.4}$ & $T_{cc}$ & $3875$ \\
\cline{2-9} & 
$D \bar{D}$ & $0$ ($0^{++}$) & $0$ & $1.38$ & ${(0.0)}^V$ & $3734.5^{+0.0(B)}_{-1.7}$ & - &
$3730.5^{+3.7}_{-5.0}$~\cite{Prelovsek:2020eiw} \\
& $D_s \bar{D}_s$ & $0$ ($0^{++}$) & $0$ & $1.00$ & ${(8.7)^V}$ & Input~\cite{Ji:2022uie} & $X(3960)$&
$3930^{+3.8}_{-2.0}$~\cite{Prelovsek:2020eiw}, 
${(3928 \pm 3)}^{V/B}$~\cite{Ji:2022uie} \\
& $D^* \bar{D}$ & $0$ ($1^{++}$) & $0$ & $1.97$ & $9.8$ & $3865.9^{+8.5}_{-17.3}$ & $X(3872)$ & $3871.7$ \\
& $D^* \bar{D}^*$ & $0$ ($2^{++}$) & $0$ & $2.04$ & $12$ & $4005^{+10}_{-19}$ & - & $4014.3 \pm 4.0 \pm 1.5$~\cite{Belle:2021nuv} \\
\cline{2-9} &
$D^* \bar{D}$ & $1$ ($1^{+}$) & $0$ & $0.88$ & ${(17.9)}^V$ & $3857.9^{+5.5}_{-6.5}$ & $Z_c(3900)$ & ${(3831-3844)}^V$~\cite{Albaladejo:2015lob} \\
& $D^* \bar{D}_s$-$D \bar{D}_s^*$ & $\frac{1}{2}$ ($1^{+}$) & $-1$ & $0.88$ & ${(15.8)}^V$ & $3962.4^{+4.8}_{-5.8}$ & $Z_{cs}(3985)$ & ${(3971-3974)^V}$~\cite{Yang:2020nrt} \\
\hline\hline
{\multirow{6}{*}{$T_{cc}(3875)$}} &
$D^* {D}$ & $0$ ($1^+$) & $0$ & $1.00$ & ${1.065}$ & Input & $T_{cc}$ &
$3874.7$ \\
\cline{2-9} & 
$D \bar{D}$ & $0$ ($0^{++}$) & $0$ & $1.57$ & ${32}$ & $3703_{-30}^{+20}$ & - &
$3730.5^{+3.7}_{-5.0}$~\cite{Prelovsek:2020eiw} \\
& $D_s \bar{D}_s$ & $0$ ($0^{++}$) & $0$ & $1.14$ & ${5.0}$ & $3931.7_{-2.7}^{+1.8}$ & $X(3960)$ & 
$3930^{+3.8}_{-2.0}$~\cite{Prelovsek:2020eiw},
${(3928 \pm 3)}^{V/B}$~\cite{Ji:2022uie} \\
& $D^* \bar{D}$ & $0$ ($1^{++}$) & $0$ & $2.23$ & $90$ & $3786_{-86}^{+70}$ & $X(3872)$ & $3871.7$ \\
& $D^* \bar{D}^*$ & $0$ ($2^{++}$) & $0$ & $2.31$ & $95$ & $3922^{+72}_{-90}$ & - &
$4014.3 \pm 4.0 \pm 1.5$~\cite{Belle:2021nuv} \\
\cline{2-9} &
$D^* \bar{D}$ & $1$ ($1^{+}$) & $0$ & $0.97$ & ${0.6}$ & $3875.2^{+0.2}_{-0.2}$ & $Z_c(3900)$ & ${(3831-3844)}^V$~\cite{Albaladejo:2015lob} \\
& $D^* \bar{D}_s$-$D \bar{D}_s^*$ & $\frac{1}{2}$ ($1^{+}$) & $-1$ & $1.00$ & $1.0$ & $3977.1^{+0.3}_{-0.2}$ & $Z_{cs}(3985)$ & ${(3971-3974)^V}$~\cite{Yang:2020nrt} \\
\hline\hline
{\multirow{6}{*}{$D\bar{D}$ (Lattice)}} &
$D^* {D}$ & $0$ ($1^+$) & $0$ & $0.64$ & ${(4.5)}^V$ & $3871.3^{+4.0}_{-9.0}$ & $T_{cc}$ & $3874.7$ \\
\cline{2-9} & 
$D \bar{D}$ & $0$ ($0^{++}$) & $0$ & $1.0$ & ${4.0}$ & $3730.5$ & - &
$3730.5^{+3.7}_{-5.0}$~\cite{Prelovsek:2020eiw} \\
& $D_s \bar{D}_s$ & $0$ ($0^{++}$) & $0$ & $0.72$ & ${(1.0)}^V$ & $3935.7^{+1.0(B)}_{-3.5}$ & $X(3960)$ & 
$3930^{+3.8}_{-2.0}$~\cite{Prelovsek:2020eiw},
${(3928 \pm 3)}^{V/B}$~\cite{Ji:2022uie} \\
& $D^* \bar{D}$ & $0$ ($1^{++}$) & $0$ & $1.42$ & $30$ & $3845 \pm 13$ & $X(3872)$ & $3871.7$ \\
& $D^* \bar{D}^*$ & $0$ ($2^{++}$) & $0$ & $1.47$ & $34$ & $3983 \pm 15$ & - &
$4014.3  \pm 4.0 \pm 1.5$~\cite{Belle:2021nuv} \\
\cline{2-9} &
$D^* \bar{D}$ & $1$ ($1^{+}$) & $0$ & $0.63$ & ${(5.4)}^V$ & $3870.4^{+4.8}_{-10.4}$ & $Z_c(3900)$ & ${(3831-3844)}^V$~\cite{Albaladejo:2015lob} \\
& $D^* \bar{D}_s$-$D \bar{D}_s^*$ & $\frac{1}{2}$ ($1^{+}$) & $-1$ & $0.64$ & ${(4.4)}^V$ & $3973.8^{+4.1}_{-9.6}$ & $Z_{cs}(3985)$ & ${(3971-3974)^V}$~\cite{Yang:2020nrt} \\
\hline\hline
\end{tabular}
\caption{Basic set of predictions for known molecular candidates observed in
  experiments or calculated in the lattice and their mutual compatibility.
  We consider three sets of predictions depending on whether we use
  the $X(3872)$, the $X(3960)$ or the $T_{cc}(3875)$ as the input
  (or reference state) for our saturation model.
  ``Input'' refers to the input state used to determine the proportionality
  constant in Eq.~(\ref{eq:coupling-sat}),
  ``System'' indicates the particular two-meson system under consideration,
  $I (J^{P(C)})$ refers to its spin, parity and C-parity (when applicable),
  $S$ to its strangeness, $R_{\rm mol}$ is the central value
  ($m_S = 475\,{\rm MeV}$) of the molecular ratio
  defined in Eq.~(\ref{eq:Rmol}), $B_{\rm mol}$ is the central value of
  the binding energy, $M_{\rm mol}$ is the mass of the molecular state
  with uncertainties (calculated as explained in Sect.~\ref{subsec:errors}),
  ``Candidate'' refers to a known resonance that might correspond to
  the two-meson system we are considering and $M_{\rm candidate}$ is
  the mass of such a candidate (this includes not only experimental
  results, but also theoretical analyses of the results and
  lattice data).
  The superscript ``V'' above the binding energy indicates a virtual state,
  while a ``$(B)$'' attached to the upper error of the mass
  indicates that the state can change from virtual to bound
  within the uncertainties of the model.
  All the binding energies and masses are in units of ${\rm MeV}$.
}
\label{tab:basic-predictions}
\end{table*}

\subsection{Tetraquark decays}

A different piece of information we might take into account
is the decay of the $T_{cc}$ into $DD\pi$ and $DD\gamma$.
The decay width of the $T_{cc}$ is expected to come mostly from the $D^* D$
component of its wave function~\cite{Yan:2021wdl}.
If we assume a wave function of the type
\begin{eqnarray}
  | T_{cc} \rangle =
  \cos{\theta}_C | D^* D \rangle +
  \sin{\theta}_C | cc \bar{u}\bar{d} \rangle \, ,
\end{eqnarray}
which contains a molecular and compact component, then the actual decay
width of the $T_{cc}$ will be dominated by the $D^* D$
components~\footnote{While the $J^P = 1^+$ $cc \bar{u} \bar{d}$ compact
  tetraquark is often predicted above the $D^* D$ threshold~\cite{Silvestre-Brac:1993zem,Semay:1994ht,Maiani:2019lpu,Luo:2017eub} and
  thus decays into this channel, this is not the case
  if the tetraquark is below the $D^* D$ threshold,
  which is what happens if the $T_{cc}^+(3875)$
  is a mixture of compact and molecular component.}.
In particular, a dimensional estimation yields~\cite{Yan:2021wdl}
\begin{eqnarray}
  \Gamma(T_{cc}) &=&
  \cos^2{\theta_C}\,\Gamma_{\rm mol} + \sin{2\theta_C}\,
  \Gamma_{\rm int} + \sin^2{\theta_C}\,\Gamma_{\rm cc}
  \nonumber \\
  &=& \cos^2{\theta_C}\,\Gamma_{\rm mol}\,\times\,
  \left[ 1 + \mathcal{O}( \tan\,{\theta_C}\,{(\frac{Q}{M_C})}^{3/2}) \right] \, , \nonumber \\
  \label{eq:P-comp}
\end{eqnarray}
where $\Gamma_{\rm mol} = \Gamma(T_{cc}(D^* D))$ and
$\Gamma_{\rm cc} = \Gamma(T_{cc}(cc \bar{u}\bar{d}))$ are the decay widths
of the molecular and compact parts of the wave function, while
$\Gamma_{\rm int}$ is an interference term (not necessarily
positive: despite the notation, it is not a decay width
per se).
In the second line we have taken into account that the decay amplitude into
$DD\,\pi$ and $DD\,\gamma$ scales as $1/Q^{3/2}$ and
$1/M_C^{3/2}$~\cite{Yan:2021wdl}, where
$Q \sim \gamma_2$ is the wave number of the $T_{cc}$ as a $D^* D$
two-body system (about $45\,{\rm MeV}$ in the isospin symmetric limit) and
$M_C$ the expected natural momentum scale at which the structure of
the compact tetraquark component is resolved.
The naive expectation is for $M_C$ to be considerably larger than
the wave number of the molecular part of the wave function, i.e.
$M_C \gg \gamma_2$.
As a consequence, in a first approximation we can ignore the compact
components of the wave function for calculating
the decay width of the $T_{cc}$: from the previous scaling,
  we expect $\Gamma_{\rm int} / \Gamma_{\rm mol} \propto (Q/M_C)^{3/2}$
  and $\Gamma_{\rm cc} / \Gamma_{\rm mol} \propto (Q/M_C)^{3}$.
  That is,
if the molecular prediction for $\Gamma_{\rm mol}$ overshots
$\Gamma(T_{cc})$ we will be able to estimate the degree of
{\it molecularness} of the $T_{cc}$.
Provided we calculate the decays of the $T_{cc}$ within pionless EFT,
the second line of Eq.~(\ref{eq:P-comp}) will be valid up to
next-to-leading order (${\rm NLO}$),
as explained in~\cite{Yan:2021wdl}.

Within the EFT formulation of~\cite{Yan:2021wdl} the ${\rm NLO}$
calculation of $\Gamma_{\rm mol}$ depends on the $T_{cc}^+$ $D^* D$
binding energy, the $D^* D$ effective range and the scattering
length of the final $DD$ two-body system.
The value of the scattering length and the effective range of a two-hadron
system in our model is given by 
\begin{eqnarray}
  \frac{1}{a_0} &=& \frac{\mu_{\rm mol}}{2 \pi C^{\rm sat}_{\rm mol}} +
  \frac{2}{\pi}\,\int_0^{\infty} dp\,f^2(\frac{p}{\Lambda}) \, ,  \\
  r_0 &=& - \frac{4}{\pi}\,\int_0^{\infty}\,\frac{dp}{p^2}\,(f^2(\frac{p}{\Lambda}) - f(0)) \nonumber \\
  && + \frac{1}{a_0}\,\frac{d^2}{d p^2}\left[ f^2(\frac{p}{\Lambda}) \right] \Big|_{p = 0}\, ,
\end{eqnarray}
where we remind that $f(x)$ refers to the regulator function.
The values for $a_0(DD)$ and $r_0(D^* D, I=0)$ can in turn be used as input
for the formulas in~\cite{Yan:2021wdl} to predict $\Gamma_{\rm mol}$.
In regards to this comparison we find:
\begin{itemize}
\item[(vii)] If the $X(3872)$ is used as input, the $DD$ scattering length
  is predicted to be $a_0 = -0.26^{+0.15}_{-0.32}\,{\rm fm}$, which translates
  into a molecular decay width of
  \begin{eqnarray}
    \Gamma_{\rm mol}^{\rm NLO} = 63.8^{+10.0}_{-7.3}\,{\rm KeV} \, ,
  \end{eqnarray}
  where the uncertainties have been calculated as in~\cite{Yan:2021wdl}
  (i.e. from the following sources: pion axial coupling, magnetic moments
  of the charmed mesons, binding energy of the $T_{cc}$ as obtained
  in the unitarized Breit-Wigner parametrization $\delta m_U$
  and the EFT truncation error),
  except for the addition in quadrature of the errors coming
  from the $DD$ scattering length.
  This is to be compared with the decay width as extracted from
  the unitarized Breit-Wigner parametrization, $\Gamma_U = 48^{+2}_{-12}
  \,{\rm KeV}$~\cite{LHCb:2021auc}, leading to
  \begin{eqnarray}
    \frac{\Gamma_U}{\Gamma_{\rm mol}^{\rm NLO}} =
    \cos^2{\theta_C} = 0.75^{+0.10}_{-0.21} \, ,
  \end{eqnarray}
  that is, this will imply that the $T_{cc}$ is about $75\%$ molecular.
  This is consistent with the fact that its location would not be
  reproduced from the molecular degrees of freedom alone,
  thus requiring contributions from other components of the wave function.
  
\item[(viii)] In contrast, if the $T_{cc}^+$ is used as input, its decay width
  happens to be too large owing to a larger $DD$ scattering length
  ($a_0 = -1.15^{+0.66}_{-1.65}\,{\rm fm}$).
  By repeating the steps in the previous comparison, we find
    \begin{eqnarray}
    \Gamma_{\rm mol}^{\rm NLO} = 85.7^{+49.5}_{-18.6}\,{\rm KeV} \, ,
    \end{eqnarray}
    which in turn leads to 
      \begin{eqnarray}
    \frac{\Gamma_U}{\Gamma_{\rm mol}^{\rm NLO}} =
    \cos^2{\theta_C} = 0.55^{+0.15}_{-0.24} \, ,
      \end{eqnarray}
      or about $55\%$ molecular.
  This implies a degree of non-molecularness that is probably not compatible
  with the initial assumption that the $T_{cc}^+$ is bound by
  the molecular components alone.
  
\end{itemize}
Thus, this comparison favors the use of the $X(3872)$ as input and a $T_{cc}^+$
state for which binding results as a combination of the interplay between
mesonic and quark degrees of freedom, as in~\cite{Janc:2004qn}.
Yet, as with the previous comparisons, caution is advised:
there is a sizable level of uncertainty associated
with all the comparisons we have made.

\begin{table*}[!ttt]
\begin{tabular}{|ccccccc|}
\hline\hline
System & $I$($J^{P(C)}$) & $R_{\rm mol}$ & $B_{\rm mol}$ & $M_{\rm mol}$
& Candidate & $M_{\rm candidate}$\\
\hline \hline
$D \bar{D}$ & $0$ ($0^{++}$) & $0.70$ & ${(1.5)}^V$ & $3733.0^{+1.2}_{-1.9}$ & - & 
$3730.5^{+3.7}_{-5.0}$~\cite{Prelovsek:2020eiw}
\\
$D^* \bar{D}$ & $0$ ($1^{++}$) & $1.00$ & ${4.1}$ & Input & $X(3872)$ & $3871.7$
\\
$D^* \bar{D}$ & $0$ ($1^{+-}$) & $0.46$ & ${(22.3)^V}$ & $3854^{+16}_{-22}$ & - &  - \\
$D^* \bar{D}^*$ & $0$ ($0^{++}$) & $0.20$ & - & - & - &  - \\
$D^* \bar{D}^*$ & $0$ ($1^{+-}$) & $0.48$ & ${(18.9)^V}$ & $3998^{+14}_{-20}$ & - & -  \\
$D^* \bar{D}^*$ & $0$ ($2^{++}$) & $1.04$ & ${5.5}$ & $4011.6 \pm 0.7$ & - & $4014.3 \pm 4.0 \pm 1.5$~\cite{Belle:2021nuv}  \\
\hline\hline
$D \bar{D}$ & $1$ ($0^{++}$) & $0.42$ & ${(30.2)}^V$ & $3704^{+21}_{-28}$ & - &  - \\
$D^* \bar{D}$ & $1$ ($1^{+\pm}$) & $0.44$ & ${(26.2)^V}$ & $3850^{+19}_{-26}$ & - &  ${(3831-3844)}^V$~\cite{Albaladejo:2015lob} \\
$D^* \bar{D}^*$ & $1$ ($0^{++}$,$1^{-+}$,$2^{++}$) & $0.46$ & ${(22.5)^V}$ & $3995^{+16}_{-24}$ & - & - \\
\hline\hline
$D \bar{D}_s$ & $\frac{1}{2}$ ($0^{+}$) & $0.43$ & ${(24.7)}^V$ & $3811^{+19}_{-28}$ & - &  - \\
$D^* \bar{D}_s$ & $\frac{1}{2}$ ($1^{+}$) & $0.45$ & ${(21.0)^V}$ & $3957^{+17}_{-26}$ & - &   ${(3971-3974)^V}$~\cite{Yang:2020nrt} \\
$D^* \bar{D}_s^*$ & $\frac{1}{2}$ ($0^{+}$,$1^{+}$,$2^{+}$) & $0.47$ & ${(17.5)^V}$ & $4103^{+14}_{-24}$ & - & -  \\
\hline\hline
$D_s \bar{D}_s$ & $0$ ($0^{++}$) & $0.51$ & ${(8.2)}^V$ & $3928.5^{+7.7}_{-15.3}$ & - & $3930^{+3.8}_{-2.0}$~\cite{Prelovsek:2020eiw}, 
       ${(3928 \pm 3)}^{V/B}$~\cite{Ji:2022uie} \\
$D_s^* \bar{D}_s$ & $0$ ($1^{++}$) & $0.64$ & ${(0.3)}^V$ & $4080.2^{+0.3(B)}_{-3.2}$ & - & - \\
$D_s^* \bar{D}_s$ & $0$ ($1^{+-}$) & $0.42$ & ${(21.9)}^V$ & $4059^{+20}_{-36}$ & - &  - \\
$D_s^* \bar{D}_s^*$ & $0$ ($0^{++}$) & $0.32$ & ${(48.1)}^V$ & $4176^{+40}_{-127}$ & - &  - \\
$D_s^* \bar{D}_s^*$ & $0$ ($1^{+-}$) & $0.44$ & ${(18.2)}^V$ & $4206^{+17}_{-34}$ & - & -  \\
$D_s^* \bar{D}_s^*$ & $0$ ($2^{++}$) & $0.66$ & ${(0.0)}^V$ & $4224.4^{+0.0(B)}_{-2.2}$ & - & -  \\
\hline\hline
\end{tabular}
\caption{
  Predictions for the $D^{(*)} \bar{D}^{(*)}$ systems
  when the $X(3872)$ is used as the reference state.
  The meaning of ``System'', ``$I(J^{P(C)})$'', ``$R_{\rm mol}$'', ``$B_{\rm mol}$'',
  ``$M_{\rm mol}$'', ``Candidate'' and ``$M_{\rm candidate}$'' is the same as
  in Table \ref{tab:basic-predictions}.
  The cutoff for the $D_s$ and $D_s^*$ charmed-strange mesons is set to
  $\Lambda_s = 1.2 \,{\rm GeV}$ (instead of $\Lambda = 1.0\,{\rm GeV}$
  in Table \ref{tab:basic-predictions}).
  All binding energies and masses are in units of ${\rm MeV}$.
}
\label{tab:predictions-hidden-X3872}
\end{table*}

\begin{table}[ht!]
\begin{tabular}{|ccccc|}
\hline\hline
System & $I(R)$($J^{P}$) & $R_{\rm mol}$ & $B_{\rm mol}$ & $M_{\rm mol}$ \\
\hline \hline
$D {D}$ & $1$ ($0^{+}$) & $0.28$ & - & - \\
$D^* {D}$ & $0$ ($1^{+}$) & $0.45$ & ${(24.2)}^V$ & $3852^{+17}_{-24}$ \\
$D^* {D}$ & $1$ ($1^{+}$) & $0.16$ & - &-  \\
$D^* {D}^*$ & $1$ ($0^{+}$) & $0.58$ & ${(7.7)}^V$ & $4009.4^{+5.8}_{-8.6}$ \\
$D^* {D}^*$ & $0$ ($1^{+}$) & $0.47$ & ${(20.7)}^V$ & $3997^{+15}_{-22}$  \\
$D^* {D}^*$ & $1$ ($2^{+}$) & $0.16$ & - & - \\
\hline\hline
$D_s {D}$ & $\tfrac{1}{2} (6)$ ($0^{+}$) & $0.34$ & ${(49.8)}^V$ & $3786^{+40}_{-63}$  \\
$D_s^* {D}$-$D_s D^*$ & $\tfrac{1}{2} (\bar{3})$ ($1^{+}$) & $0.43$ & ${(25.0)}^V$ & $3953^{+20}_{-31}$  \\
$D_s^* {D}$-$D_s D^*$ & $\tfrac{1}{2} (6)$ ($1^{+}$) & $0.23$ & $-$ & $-$  \\
$D_s^* {D}^*$ & $\tfrac{1}{2} (6)$ ($0^{+}$) & $0.59$ & ${(4.4)}^V$ & $4116.3^{+4.0}_{-7.1}$ \\
$D_s^* {D}^*$ & $\tfrac{1}{2} (\bar{3})$ ($1^{+}$) & $0.43$ & ${(24.1)}^V$ & $4097^{+19}_{-30}$ \\
$D_s^* {D}^*$ & $\tfrac{1}{2} (6)$ ($2^{+}$) & $0.23$ & - & -  \\
\hline \hline
$D_s {D}_s$ & $0$ ($0^{+}$) & $0.38$ & ${(32.7)}^V$ & $3804^{+28}_{-50}$ \\
$D_s^* {D}_s$ & $0$ ($1^{+}$) & $0.28$ & - & -  \\
$D_s^* {D}_s^*$ & $0$ ($0^{+}$) & $0.63$ & ${(0.4)}^V$ & $4224.0^{+0.4(B)}_{-4.0}$  \\
$D_s^* {D}_s^*$ & $0$ ($2^{+}$) & $0.29$ & - & -  \\
\hline \hline
\end{tabular}
\caption{
  Predictions for the $D^{(*)} {D}^{(*)}$ systems
  when the $X(3872)$ is used as the reference state.
  The meaning of ``System'', ``$I(R)(J^{P})$'', ``$R_{\rm mol}$'', ``$B_{\rm mol}$''
  and ``$M_{\rm mol}$'' is the same as in Table \ref{tab:basic-predictions},
  where the only difference is that now after isospin we include ``$(R)$''
  in parentheses to indicate the SU(3)-flavor representation to which
  a $D_s^{(*)} D^{(*)}$ state belong (the reason being that they are not
  distinguishable by isospin alone in this case).
  We do not include candidate states and their masses as
  there is only one: the $T_{cc}^+(3875)$ with a mass of
  $3875.7\,{\rm MeV}$.
  The cutoff for the $D_s$ and $D_s^*$ charmed-strange mesons is set to
  $\Lambda_s = 1.2 \,{\rm GeV}$ (instead of $\Lambda = 1.0\,{\rm GeV}$
  in Table \ref{tab:basic-predictions}).
  All binding energies and masses are in units of ${\rm MeV}$.
}
\label{tab:predictions-open-X3872}
\end{table}

\begin{table}[!ttt]
\begin{tabular}{|ccccc|}
\hline\hline
System & $I(R)$($J^{P}$) & $R_{\rm mol}$ & $B_{\rm mol}$ & $M_{\rm mol}$ \\
\hline \hline
$D {D}$ & $1$ ($0^{+}$) & $0.63$ & ${(9.9)}^V$ & $3724.6^{+7.9}_{-24.8}$ \\
$D^* {D}$ & $0$ ($1^{+}$) & $1.0$ & ${1.065}$ & $3874.7$ \\
$D^* {D}$ & $1$ ($1^{+}$) & $0.35$ & - &-  \\
$D^* {D}^*$ & $1$ ($0^{+}$) & $1.30$ & ${12.2}$ & $4004.9^{+5.6}_{-8.5}$ \\
$D^* {D}^*$ & $0$ ($1^{+}$) & $1.04$ & $1.8$ & $4015.3 \pm 0.1$  \\
$D^* {D}^*$ & $1$ ($2^{+}$) & $0.36$ & - & - \\
\hline\hline
$D_s {D}$ & $\tfrac{1}{2} (6)$ ($0^{+}$) & $0.75$ & ${(0.8)}^V$ & $3834.8^{+0.8(B)}_{-6.1}$ \\
$D_s^* {D}$-$D_s D^*$ & $\tfrac{1}{2} (\bar{3})$ ($1^{+}$) & $0.95$ & ${2.1}$ & $3976.1^{+1.1}_{-1.0}$ \\
$D_s^* {D}$-$D_s D^*$ & $\tfrac{1}{2} (6)$ ($1^{+}$) & $0.51$ & ${(19.2)}^V$ & $3959^{+18}_{-90}$ \\
$D_s^* {D}^*$ & $\tfrac{1}{2} (6)$ ($0^{+}$) & $1.32$ & $23.1$ & $4098^{+11}_{-15}$ \\
$D_s^* {D}^*$ & $\tfrac{1}{2} (\bar{3})$ ($1^{+}$) & $0.95$ & ${2.0}$ & $4188.7 \pm 1.0$ \\
$D_s^* {D}^*$ & $\tfrac{1}{2} (6)$ ($2^{+}$) & $0.51$ & ${(18.5)}^V$ & $4102^{+17}_{-87}$ \\
\hline \hline
$D_s {D}_s$ & $0$ ($0^{+}$) & $0.85$ & ${2.0}$ & $3934.7^{+1.9}_{-3.2}$ \\
$D_s^* {D}_s$ & $0$ ($1^{+}$) & $0.63$ & ${(2.8)}^V$ & $4077.8^{+2.8(B)}_{-31.0}$ \\
$D_s^* {D}_s^*$ & $0$ ($0^{+}$) & $1.41$ & ${52.8}$ & $4172^{+28}_{-32}$ \\
$D_s^* {D}_s^*$ & $0$ ($2^{+}$) & $0.66$ & ${(1.6)}^V$ & $4222.8^{+1.6(B)}_{-27.4}$ \\
\hline \hline
\end{tabular}
\caption{
  Predictions for the $D^{(*)} {D}^{(*)}$ systems
  when the $T_{cc}^+$ is used as the reference state.
  The conventions used are identical to those of
  Table \ref{tab:predictions-open-X3872}.
}
\label{tab:predictions-open-Tcc3875}
\end{table}

\subsection{Choice of the reference state}

All in all, from a comparison of the (partially known) spectrum of two charmed
meson systems, the expected interference from non-molecular degrees of freedom
and the decays of the $T_{cc}^+$, it seems that using the $X(3872)$ as input
is the better choice.

This is not to say that the $X(3872)$ is purely molecular, only that
the assumption that its binding can be purely explained in terms of
its molecular components is more congruent with the previous
information than the analogous assumption for the $T_{cc}^+$.
Besides, there are certain aspects of the $X(3872)$ that are better
described by assuming the existence of short-range compact
components~\cite{Takizawa:2012hy,Ferretti:2014xqa,Guo:2014taa}.
Also, it has been argued that the effective range for the $D^*\bar{D}$ pair
within the $X(3872)$ is negative~\cite{Esposito:2021vhu},
which might require contributions from degrees of freedom different
than $D^*\bar{D}$ to the scattering amplitude.
Nonetheless we consider the $X(3872)$ to be the best choice for the reference
state and compute the molecular spectrum from it.
Yet, for the particular case of the $D^{(*)} D^{(*)}$ two-body system,
we will consider too the predictions when the $T_{cc}^+$ is used
as the reference state. This will yield a different set of
molecular states that, if eventually observed,
could be used to decide between the $X(3872)$ and $T_{cc}^+$
as the most molecular of the two.

\section{Predictions}
\label{sec:predictions}

For the predictions of the molecular spectrum of two charmed mesons
we will use the $X(3872)$ as the reference state,
as previously explained.
This leads to $C^{\rm sat}_{\rm ref} = -0.79\,{\rm fm}^2$
for the coupling of the reference state.

We will also include a modification to our model
when strange charmed mesons are included.
This modification is as follows: instead of using our original formulation
in Eq.~(\ref{eq:Vc-reg}), which implicitly assumes the same cutoff
for the two hadrons within a molecular state, we will consider
the possibility that each of the hadrons is better described
by a different cutoff
\begin{eqnarray}
  V_C = C^{\rm sat} f(\frac{p'}{\Lambda_1}) f(\frac{p}{\Lambda_2}) \, .
  \label{eq:Vc-reg12}
\end{eqnarray}
In particular we will distinguish between the non-strange and strange sectors:
for the $D$ and $D^*$ charmed mesons we will use $\Lambda = 1.0\,{\rm GeV}$,
just as before, but for $D_s$ and $D_s^*$ we will use 
$\Lambda = 1.2\,{\rm GeV}$ instead.
The reasons for this change are that it takes into account the more compact
nature of the strange-charmed mesons and that it help us reproduce 
the location of the virtual state interpretation of the $X(3960)$:
with $\Lambda = 1.2\,{\rm GeV}$ we obtain $3927.2\,{\rm MeV}$,
which is compatible with the $3928 \pm 3\,{\rm MeV}$ pole
mass of~\cite{Ji:2022uie}.

With the previous choices, in the hidden charmed sector we obtain 
the spectrum we show in Table~\ref{tab:predictions-hidden-X3872}.
Only two systems clearly bind below threshold, the $1^{++}$ $D^* \bar{D}$
and $2^{++} D^* \bar{D}^*$ configurations, yet there are a few other
systems that are very close to binding: $0^{++}$ $D\bar{D}$,
$1^{++}$ $D_s^* \bar{D}_s$ and $2^{++}$ $D_s^* \bar{D}_s^*$.

For the doubly charmed sector we have made two sets of prediction, depending
on whether we use the $X(3872)$ or the $T_{cc}^+$ as input, Tables  
\ref{tab:predictions-open-X3872} and \ref{tab:predictions-open-Tcc3875},
respectively.
The doubly charmed molecules are constrained by (extended)
Bose-Einstein symmetry, i.e. not all combinations of
spin and isospin are allowed.
For $D D$ and $D^* D^*$ this translates into the condition that $(I+J)$ must
be an odd number, while for $D_s D_s$ and $D_s^* D_s^*$ the condition is
that $J$ must be even.
For the $D_s^{(*)} D^{(*)}$ there are two configurations, antitriplet and sextet,
given by
\begin{eqnarray}
  | \bar{3} \rangle &=& \frac{1}{\sqrt{2}}\,\left[ |D_s^{(*)} D^{(*)} \rangle -
    |D^{(*)} D_s^{(*)} \rangle \right] \, , \\
  | 6 \rangle &=& \frac{1}{\sqrt{2}}\,\left[ |D_s^{(*)} D^{(*)} \rangle +
    |D^{(*)} D_s^{(*)} \rangle \right] \, .
\end{eqnarray}
The antitriplet and sextet configurations correspond to the SU(3)-flavor
generalizations of the $I=0$ and $I=1$ configurations in $D^{(*)} D^{(*)}$.
Indeed, it can be readily noticed in Tables \ref{tab:predictions-open-X3872}
and \ref{tab:predictions-open-Tcc3875} that the binding or virtual state
energies of the $I=0$ ($I=1$) $D^{(*)} D^{(*)}$ and $\bar{3}$ ($6$)
$D_s^{(*)} D^{(*)}$ molecules is predicted to be similar.
This consequence of SU(3)-flavor symmetry was previously exploited in~\cite{Dai:2021vgf} to predict the $D_s D^*$-$D_s^* D$ molecular partners of the $T_{cc}$.

When we use the $X(3872)$ as input, Table \ref{tab:predictions-open-X3872},
it is apparent that the $D^* D^*$ HQSS partner of the $T_{cc}$,
i.e. the $T_{cc}^*$, does not bind and it is a virtual state
not close to threshold instead.
This is in contrast with the prediction of a bound $T_{cc}^*$ when the $T_{cc}^+$
is used as input, Table \ref{tab:predictions-open-Tcc3875}.
This difference in the predictions could be used to better understand
the nature of the $T_{cc}^+$: the eventual discovery of the $T_{cc}^*$
at about $(1-2)\,{\rm MeV}$ below the $D^* D^*$ threshold would be
a strong indication that the $T_{cc}^+$ is mostly molecular.
If it were not and the binding of the $T_{cc}^+$ required the interplay between
the mesonic and quark degrees of freedom, we would not expect
the $T_{cc}^*$ to bind: if the pole of the compact component
once we remove the coupling with the $D^* D$ channel were
to be between the $D^* D$ and $D^* D^*$ threshold,
this component will favor binding for $D^* D$ and
disfavor the formation of a $D^* D^*$ molecule.
However, this is not necessarily the only possibility: were this isolated
compact component to be somewhat above the $D^* D^*$ threshold, the HQSS
expectation that the $J=1$ $D^* D$ and $D^* D^*$ states have the same
binding energy would be strongly violated.
Yet, this later possibility seems less plausible as it would require a really
strong coupling between the compact and molecular components for providing
strong enough attraction in the $D^* D$ channel.

Be it as it may, this is not the only difference between the two spectra.
The doubly charmed spectrum derived from the $X(3872)$ contains very few
configurations close to binding, basically the sextet $J^P = 0^+$
configurations containing two excited charmed mesons,
i.e. the $I=1$ $J^P=0^+$ $D^* D^*$, the sextet $0^+$ $D^* D_s^*$ and
the $0^+$ $D_s^* D_s^*$ molecules.
In contrast, if the $T_{cc}^+$ were to be mostly molecular, the spectrum of
possible molecular states would be much richer.
For instance, all the antitriplet $J^P = 1^+$ configurations will bind,
including the $T_{cc}$ and $T_{cc}^*$ as well as their strange and
hidden-strange counterparts.
Again, the eventual detection of these molecules
will imply a molecular $T_{cc}^+$.
Alternatively, lattice calculations of the $D_s^{(*)} D^{(*)}$ and
$D_s^{(*)} D_s^{(*)}$ systems will shed light on this issue,
though for the moment only calculations in the $D^* D$ case exist
(indicating either a virtual~\cite{Prelovsek:2020eiw,Padmanath:2022cvl}
or bound state~\cite{Lyu:2023xro} solution for the $T_{cc}^+$).
Yet, independently of the input state, the most attractive configuration
turns out to be the $I=0$ or sextet $J^P = 0^+$ $D^* D^*$ and
$0^+$ $D^* D_s^*$ molecules, a conclusion which is in agreement
with~\cite{Ke:2021rxd} for the non-strange sector.

\section{Conclusions}
\label{sec:conclusion}

To summarize, we have considered the molecular spectrum of systems containing
two S-wave charmed mesons within a contact-range theory in which
the couplings are saturated by light-meson exchanges
($\sigma$, $\rho$, $\omega$).
The question we wanted to address was whether the $X(3872)$, $X(3960)$ and
$T_{cc}^+(3875)$ can all be described with the same set of parameters.
It turns out that this is not the case and that there is a tension between
the molecular description of the $X(3872)$ and $T_{cc}^+(3875)$
within the saturation model we use.
Basically, if the $X(3872)$ is molecular it would be difficult to explain
the $T_{cc}^+(3875)$ in purely molecular terms, and vice versa.

Intuitively this can be understood in terms of vector meson exchange alone,
as the attraction provided by this effect is twice as big
in the isoscalar $D^*\bar{D}$ system than
in the $D^* {D}$ one
\begin{eqnarray}
  && V_V(D^{*}\bar{D},I=0) = \nonumber \\ && \quad
  - 4\,\left[ \frac{g_V^2}{m_V^2 + {\vec{q}\,}^2} +
    \frac{f_V^2}{6 M^2}\,\frac{m_V^2}{m_V^2 + {\vec{q}\,}^2} \,
    \vec{\sigma}_{L1} \cdot \vec{\sigma}_{L2}
    \right] \, , \\
  && V_V(D^{*}{D},I=0) = \nonumber \\ && \quad
  - 2\,\left[ \frac{g_V^2}{m_V^2 + {\vec{q}\,}^2} +
    \frac{f_V^2}{6 M^2}\,\frac{m_V^2}{m_V^2 + {\vec{q}\,}^2} \,
    \vec{\sigma}_{L1} \cdot \vec{\sigma}_{L2}
    \right] \, , 
\end{eqnarray}
as derived from Eq.~(\ref{eq:V-V}) once we include isospin factors or
as in Eqs.~(\ref{eq:C0-V}) and (\ref{eq:C1-V}) once we use
the contact-range approximation.
Thus, in the absence of other attractive effects, we expect
a molecular $X(3872)$ to be considerably more bound than
a molecular $T_{cc}^+(3875)$.
The inclusion of the scalar meson, which provides the same degree of attraction
in both systems, somewhat softens the previous conclusion but not necessarily
as much as to avoid the tension (unless scalar meson exchange
is much stronger than expected here).

From a comparison of the hidden- and open-charm two-meson molecules with
known molecular candidates and the decays of the $T_{cc}^+(3875)$,
we consider that it is more probable for the $X(3872)$ to be
mostly molecular than for the $T_{cc}^+(3875)$.
Hence, to make predictions of the two charmed meson molecular spectrum
we use the $X(3872)$ as a reference or input state
within our RG-saturation model.
For the hidden charm sector, this choice generates only a second molecular
state that is clearly below threshold: the $J^{PC} = 2^{++}$ partner of the
$X(3872)$.
Yet, there are a few virtual states that are extremely close to threshold and
can bind within the uncertainties of our model.
These include the $0^{++}$ $D \bar{D}$ and $D_s \bar{D}_s$ molecules, the first
one corresponding with the state found in the lattice~\cite{Prelovsek:2020eiw}
and the second one with the $X(3960)$.

For the open charm sector, the most attractive configurations are the $I=1$,
$J=0^+$ $D^* D^*$, sextet $J=0^+$ $D_s^* D^*$ and $J=0^+$ $D_s^* D_s^*$
molecules, which are predicted as virtual states really
close to threshold.
However, the interesting feature of the open charm molecules is
that their spectrum will allow us to distinguish whether
the $T_{cc}^+(3875)$ is only partly or
predominantly molecular.
In the second case --- a $T_{cc}^+(3875)$ whose binding can be explained purely
in molecular terms --- there will be a $I=0$, $J=1^+$ $D^* D^*$ partner state
at $4015\,{\rm MeV}$, a $T_{cc}^{*+} (4015)$.
The existence of this state has been consistently predicted in models assuming
that the $T_{cc}^+(3875)$ is predominantly molecular in the first
place~\cite{Albaladejo:2021vln,Du:2021zzh,Jia:2022qwr}.
Finding the $T_{cc}^{*+} (4015)$ will thus represent a very strong hint
that the $T_{cc}^+(3875)$ binding comes almost exclusively from its
molecular components.

\section*{Acknowledgments}

This work is partly supported by the National Natural Science Foundation
of China under Grants No. 11735003, No. 11835015, No. 11975041, No. 12047503
and No. 12125507, the Chinese Academy of Sciences under Grant No. XDB34030000,
the China Postdoctoral Science Foundation under Grant No. 2022M713229,
the Fundamental Research Funds for the Central Universities and
the Thousand Talents Plan for Young Professionals.
M.P.V. would also like to thank the IJCLab of Orsay, where part of
this work has been done, for its long-term hospitality.


\begin{thebibliography}{85}%
\makeatletter
\providecommand \@ifxundefined [1]{%
 \@ifx{#1\undefined}
}%
\providecommand \@ifnum [1]{%
 \ifnum #1\expandafter \@firstoftwo
 \else \expandafter \@secondoftwo
 \fi
}%
\providecommand \@ifx [1]{%
 \ifx #1\expandafter \@firstoftwo
 \else \expandafter \@secondoftwo
 \fi
}%
\providecommand \natexlab [1]{#1}%
\providecommand \enquote  [1]{``#1''}%
\providecommand \bibnamefont  [1]{#1}%
\providecommand \bibfnamefont [1]{#1}%
\providecommand \citenamefont [1]{#1}%
\providecommand \href@noop [0]{\@secondoftwo}%
\providecommand \href [0]{\begingroup \@sanitize@url \@href}%
\providecommand \@href[1]{\@@startlink{#1}\@@href}%
\providecommand \@@href[1]{\endgroup#1\@@endlink}%
\providecommand \@sanitize@url [0]{\catcode `\\12\catcode `\$12\catcode
  `\&12\catcode `\#12\catcode `\^12\catcode `\_12\catcode `\%12\relax}%
\providecommand \@@startlink[1]{}%
\providecommand \@@endlink[0]{}%
\providecommand \url  [0]{\begingroup\@sanitize@url \@url }%
\providecommand \@url [1]{\endgroup\@href {#1}{\urlprefix }}%
\providecommand \urlprefix  [0]{URL }%
\providecommand \Eprint [0]{\href }%
\providecommand \doibase [0]{http://dx.doi.org/}%
\providecommand \selectlanguage [0]{\@gobble}%
\providecommand \bibinfo  [0]{\@secondoftwo}%
\providecommand \bibfield  [0]{\@secondoftwo}%
\providecommand \translation [1]{[#1]}%
\providecommand \BibitemOpen [0]{}%
\providecommand \bibitemStop [0]{}%
\providecommand \bibitemNoStop [0]{.\EOS\space}%
\providecommand \EOS [0]{\spacefactor3000\relax}%
\providecommand \BibitemShut  [1]{\csname bibitem#1\endcsname}%
\let\auto@bib@innerbib\@empty
\bibitem [{\citenamefont {Aaij}\ \emph {et~al.}(2023)\citenamefont {Aaij} \emph
  {et~al.}}]{LHCb:2022vsv}%
  \BibitemOpen
  \bibfield  {author} {\bibinfo {author} {\bibfnamefont {R.}~\bibnamefont
  {Aaij}} \emph {et~al.} (\bibinfo {collaboration} {LHCb}),\ }\href {\doibase
  10.1103/PhysRevLett.131.071901} {\bibfield  {journal} {\bibinfo  {journal}
  {Phys. Rev. Lett.}\ }\textbf {\bibinfo {volume} {131}},\ \bibinfo {pages}
  {071901} (\bibinfo {year} {2023})},\ \Eprint
  {http://arxiv.org/abs/2210.15153} {arXiv:2210.15153 [hep-ex]} \BibitemShut
  {NoStop}%
\bibitem [{\citenamefont {Aaij}\ \emph {et~al.}(2020)\citenamefont {Aaij} \emph
  {et~al.}}]{LHCb:2020pxc}%
  \BibitemOpen
  \bibfield  {author} {\bibinfo {author} {\bibfnamefont {R.}~\bibnamefont
  {Aaij}} \emph {et~al.} (\bibinfo {collaboration} {LHCb}),\ }\href {\doibase
  10.1103/PhysRevD.102.112003} {\bibfield  {journal} {\bibinfo  {journal}
  {Phys. Rev. D}\ }\textbf {\bibinfo {volume} {102}},\ \bibinfo {pages}
  {112003} (\bibinfo {year} {2020})},\ \Eprint
  {http://arxiv.org/abs/2009.00026} {arXiv:2009.00026 [hep-ex]} \BibitemShut
  {NoStop}%
\bibitem [{\citenamefont {Workman}(2022)}]{Workman:2022ynf}%
  \BibitemOpen
  \bibfield  {author} {\bibinfo {author} {\bibfnamefont {R.~L.}\ \bibnamefont
  {Workman}} (\bibinfo {collaboration} {Particle Data Group}),\ }\href
  {\doibase 10.1093/ptep/ptac097} {\bibfield  {journal} {\bibinfo  {journal}
  {PTEP}\ }\textbf {\bibinfo {volume} {2022}},\ \bibinfo {pages} {083C01}
  (\bibinfo {year} {2022})}\BibitemShut {NoStop}%
\bibitem [{\citenamefont {Ji}\ \emph {et~al.}(2023)\citenamefont {Ji},
  \citenamefont {Dong}, \citenamefont {Albaladejo}, \citenamefont {Du},
  \citenamefont {Guo}, \citenamefont {Nieves},\ and\ \citenamefont
  {Zou}}]{Ji:2022vdj}%
  \BibitemOpen
  \bibfield  {author} {\bibinfo {author} {\bibfnamefont {T.}~\bibnamefont
  {Ji}}, \bibinfo {author} {\bibfnamefont {X.-K.}\ \bibnamefont {Dong}},
  \bibinfo {author} {\bibfnamefont {M.}~\bibnamefont {Albaladejo}}, \bibinfo
  {author} {\bibfnamefont {M.-L.}\ \bibnamefont {Du}}, \bibinfo {author}
  {\bibfnamefont {F.-K.}\ \bibnamefont {Guo}}, \bibinfo {author} {\bibfnamefont
  {J.}~\bibnamefont {Nieves}}, \ and\ \bibinfo {author} {\bibfnamefont {B.-S.}\
  \bibnamefont {Zou}},\ }\href {\doibase 10.1016/j.scib.2023.02.034} {\bibfield
   {journal} {\bibinfo  {journal} {Sci. Bull.}\ }\textbf {\bibinfo {volume}
  {68}},\ \bibinfo {pages} {688} (\bibinfo {year} {2023})},\ \Eprint
  {http://arxiv.org/abs/2212.00631} {arXiv:2212.00631 [hep-ph]} \BibitemShut
  {NoStop}%
\bibitem [{\citenamefont {Abreu}\ \emph {et~al.}(2023)\citenamefont {Abreu},
  \citenamefont {Albaladejo}, \citenamefont {Feijoo}, \citenamefont {Oset},\
  and\ \citenamefont {Nieves}}]{Abreu:2023rye}%
  \BibitemOpen
  \bibfield  {author} {\bibinfo {author} {\bibfnamefont {L.~M.}\ \bibnamefont
  {Abreu}}, \bibinfo {author} {\bibfnamefont {M.}~\bibnamefont {Albaladejo}},
  \bibinfo {author} {\bibfnamefont {A.}~\bibnamefont {Feijoo}}, \bibinfo
  {author} {\bibfnamefont {E.}~\bibnamefont {Oset}}, \ and\ \bibinfo {author}
  {\bibfnamefont {J.}~\bibnamefont {Nieves}},\ }\href {\doibase
  10.1140/epjc/s10052-023-11467-1} {\bibfield  {journal} {\bibinfo  {journal}
  {Eur. Phys. J. C}\ }\textbf {\bibinfo {volume} {83}},\ \bibinfo {pages} {309}
  (\bibinfo {year} {2023})},\ \Eprint {http://arxiv.org/abs/2302.08877}
  {arXiv:2302.08877 [hep-ph]} \BibitemShut {NoStop}%
\bibitem [{\citenamefont {Ji}\ \emph {et~al.}(2022)\citenamefont {Ji},
  \citenamefont {Dong}, \citenamefont {Albaladejo}, \citenamefont {Du},
  \citenamefont {Guo},\ and\ \citenamefont {Nieves}}]{Ji:2022uie}%
  \BibitemOpen
  \bibfield  {author} {\bibinfo {author} {\bibfnamefont {T.}~\bibnamefont
  {Ji}}, \bibinfo {author} {\bibfnamefont {X.-K.}\ \bibnamefont {Dong}},
  \bibinfo {author} {\bibfnamefont {M.}~\bibnamefont {Albaladejo}}, \bibinfo
  {author} {\bibfnamefont {M.-L.}\ \bibnamefont {Du}}, \bibinfo {author}
  {\bibfnamefont {F.-K.}\ \bibnamefont {Guo}}, \ and\ \bibinfo {author}
  {\bibfnamefont {J.}~\bibnamefont {Nieves}},\ }\href {\doibase
  10.1103/PhysRevD.106.094002} {\bibfield  {journal} {\bibinfo  {journal}
  {Phys. Rev. D}\ }\textbf {\bibinfo {volume} {106}},\ \bibinfo {pages}
  {094002} (\bibinfo {year} {2022})},\ \Eprint
  {http://arxiv.org/abs/2207.08563} {arXiv:2207.08563 [hep-ph]} \BibitemShut
  {NoStop}%
\bibitem [{\citenamefont {Xin}\ \emph {et~al.}(2022)\citenamefont {Xin},
  \citenamefont {Wang},\ and\ \citenamefont {Yang}}]{Xin:2022bzt}%
  \BibitemOpen
  \bibfield  {author} {\bibinfo {author} {\bibfnamefont {Q.}~\bibnamefont
  {Xin}}, \bibinfo {author} {\bibfnamefont {Z.-G.}\ \bibnamefont {Wang}}, \
  and\ \bibinfo {author} {\bibfnamefont {X.-S.}\ \bibnamefont {Yang}},\ }\href
  {\doibase 10.1007/s43673-022-00070-3} {\bibfield  {journal} {\bibinfo
  {journal} {AAPPS Bull.}\ }\textbf {\bibinfo {volume} {32}},\ \bibinfo {pages}
  {37} (\bibinfo {year} {2022})},\ \Eprint {http://arxiv.org/abs/2207.09910}
  {arXiv:2207.09910 [hep-ph]} \BibitemShut {NoStop}%
\bibitem [{\citenamefont {Xie}\ \emph {et~al.}(2023)\citenamefont {Xie},
  \citenamefont {Liu},\ and\ \citenamefont {Geng}}]{Xie:2022lyw}%
  \BibitemOpen
  \bibfield  {author} {\bibinfo {author} {\bibfnamefont {J.-M.}\ \bibnamefont
  {Xie}}, \bibinfo {author} {\bibfnamefont {M.-Z.}\ \bibnamefont {Liu}}, \ and\
  \bibinfo {author} {\bibfnamefont {L.-S.}\ \bibnamefont {Geng}},\ }\href
  {\doibase 10.1103/PhysRevD.107.016003} {\bibfield  {journal} {\bibinfo
  {journal} {Phys. Rev. D}\ }\textbf {\bibinfo {volume} {107}},\ \bibinfo
  {pages} {016003} (\bibinfo {year} {2023})},\ \Eprint
  {http://arxiv.org/abs/2207.12178} {arXiv:2207.12178 [hep-ph]} \BibitemShut
  {NoStop}%
\bibitem [{\citenamefont {Mutuk}(2022)}]{Mutuk:2022ckn}%
  \BibitemOpen
  \bibfield  {author} {\bibinfo {author} {\bibfnamefont {H.}~\bibnamefont
  {Mutuk}},\ }\href {\doibase 10.1140/epjc/s10052-022-11120-3} {\bibfield
  {journal} {\bibinfo  {journal} {Eur. Phys. J. C}\ }\textbf {\bibinfo {volume}
  {82}},\ \bibinfo {pages} {1142} (\bibinfo {year} {2022})},\ \Eprint
  {http://arxiv.org/abs/2211.14836} {arXiv:2211.14836 [hep-ph]} \BibitemShut
  {NoStop}%
\bibitem [{\citenamefont {Chen}\ \emph {et~al.}(2023)\citenamefont {Chen},
  \citenamefont {Chen}, \citenamefont {Meng}, \citenamefont {Qi},\ and\
  \citenamefont {Zheng}}]{Chen:2023eix}%
  \BibitemOpen
  \bibfield  {author} {\bibinfo {author} {\bibfnamefont {Y.}~\bibnamefont
  {Chen}}, \bibinfo {author} {\bibfnamefont {H.}~\bibnamefont {Chen}}, \bibinfo
  {author} {\bibfnamefont {C.}~\bibnamefont {Meng}}, \bibinfo {author}
  {\bibfnamefont {H.-R.}\ \bibnamefont {Qi}}, \ and\ \bibinfo {author}
  {\bibfnamefont {H.-Q.}\ \bibnamefont {Zheng}},\ }\href {\doibase
  10.1140/epjc/s10052-023-11527-6} {\bibfield  {journal} {\bibinfo  {journal}
  {Eur. Phys. J. C}\ }\textbf {\bibinfo {volume} {83}},\ \bibinfo {pages} {381}
  (\bibinfo {year} {2023})},\ \Eprint {http://arxiv.org/abs/2302.06278}
  {arXiv:2302.06278 [hep-ph]} \BibitemShut {NoStop}%
\bibitem [{\citenamefont {Tornqvist}(2003)}]{Tornqvist:2003na}%
  \BibitemOpen
  \bibfield  {author} {\bibinfo {author} {\bibfnamefont {N.~A.}\ \bibnamefont
  {Tornqvist}},\ }\href@noop {} {\  (\bibinfo {year} {2003})},\ \Eprint
  {http://arxiv.org/abs/hep-ph/0308277} {arXiv:hep-ph/0308277 [hep-ph]}
  \BibitemShut {NoStop}%
\bibitem [{\citenamefont {Voloshin}(2004)}]{Voloshin:2003nt}%
  \BibitemOpen
  \bibfield  {author} {\bibinfo {author} {\bibfnamefont {M.~B.}\ \bibnamefont
  {Voloshin}},\ }\href {\doibase 10.1016/j.physletb.2003.11.014} {\bibfield
  {journal} {\bibinfo  {journal} {Phys. Lett.}\ }\textbf {\bibinfo {volume}
  {B579}},\ \bibinfo {pages} {316} (\bibinfo {year} {2004})},\ \Eprint
  {http://arxiv.org/abs/hep-ph/0309307} {arXiv:hep-ph/0309307 [hep-ph]}
  \BibitemShut {NoStop}%
\bibitem [{\citenamefont {Braaten}\ and\ \citenamefont
  {Kusunoki}(2004)}]{Braaten:2003he}%
  \BibitemOpen
  \bibfield  {author} {\bibinfo {author} {\bibfnamefont {E.}~\bibnamefont
  {Braaten}}\ and\ \bibinfo {author} {\bibfnamefont {M.}~\bibnamefont
  {Kusunoki}},\ }\href {\doibase 10.1103/PhysRevD.69.074005} {\bibfield
  {journal} {\bibinfo  {journal} {Phys. Rev.}\ }\textbf {\bibinfo {volume}
  {D69}},\ \bibinfo {pages} {074005} (\bibinfo {year} {2004})},\ \Eprint
  {http://arxiv.org/abs/hep-ph/0311147} {arXiv:hep-ph/0311147 [hep-ph]}
  \BibitemShut {NoStop}%
\bibitem [{\citenamefont {Swanson}(2004)}]{Swanson:2004pp}%
  \BibitemOpen
  \bibfield  {author} {\bibinfo {author} {\bibfnamefont {E.~S.}\ \bibnamefont
  {Swanson}},\ }\href {\doibase 10.1016/j.physletb.2004.07.059} {\bibfield
  {journal} {\bibinfo  {journal} {Phys. Lett.}\ }\textbf {\bibinfo {volume}
  {B598}},\ \bibinfo {pages} {197} (\bibinfo {year} {2004})},\ \Eprint
  {http://arxiv.org/abs/hep-ph/0406080} {arXiv:hep-ph/0406080 [hep-ph]}
  \BibitemShut {NoStop}%
\bibitem [{\citenamefont {Dong}\ \emph {et~al.}(2011)\citenamefont {Dong},
  \citenamefont {Faessler}, \citenamefont {Gutsche},\ and\ \citenamefont
  {Lyubovitskij}}]{Dong:2009uf}%
  \BibitemOpen
  \bibfield  {author} {\bibinfo {author} {\bibfnamefont {Y.}~\bibnamefont
  {Dong}}, \bibinfo {author} {\bibfnamefont {A.}~\bibnamefont {Faessler}},
  \bibinfo {author} {\bibfnamefont {T.}~\bibnamefont {Gutsche}}, \ and\
  \bibinfo {author} {\bibfnamefont {V.~E.}\ \bibnamefont {Lyubovitskij}},\
  }\href {\doibase 10.1088/0954-3899/38/1/015001} {\bibfield  {journal}
  {\bibinfo  {journal} {J. Phys.}\ }\textbf {\bibinfo {volume} {G38}},\
  \bibinfo {pages} {015001} (\bibinfo {year} {2011})},\ \Eprint
  {http://arxiv.org/abs/0909.0380} {arXiv:0909.0380 [hep-ph]} \BibitemShut
  {NoStop}%
\bibitem [{\citenamefont {Gamermann}\ and\ \citenamefont
  {Oset}(2009)}]{Gamermann:2009fv}%
  \BibitemOpen
  \bibfield  {author} {\bibinfo {author} {\bibfnamefont {D.}~\bibnamefont
  {Gamermann}}\ and\ \bibinfo {author} {\bibfnamefont {E.}~\bibnamefont
  {Oset}},\ }\href {\doibase 10.1103/PhysRevD.80.014003} {\bibfield  {journal}
  {\bibinfo  {journal} {Phys. Rev.}\ }\textbf {\bibinfo {volume} {D80}},\
  \bibinfo {pages} {014003} (\bibinfo {year} {2009})},\ \Eprint
  {http://arxiv.org/abs/0905.0402} {arXiv:0905.0402 [hep-ph]} \BibitemShut
  {NoStop}%
\bibitem [{\citenamefont {Gamermann}\ \emph {et~al.}(2010)\citenamefont
  {Gamermann}, \citenamefont {Nieves}, \citenamefont {Oset},\ and\
  \citenamefont {Ruiz~Arriola}}]{Gamermann:2009uq}%
  \BibitemOpen
  \bibfield  {author} {\bibinfo {author} {\bibfnamefont {D.}~\bibnamefont
  {Gamermann}}, \bibinfo {author} {\bibfnamefont {J.}~\bibnamefont {Nieves}},
  \bibinfo {author} {\bibfnamefont {E.}~\bibnamefont {Oset}}, \ and\ \bibinfo
  {author} {\bibfnamefont {E.}~\bibnamefont {Ruiz~Arriola}},\ }\href {\doibase
  10.1103/PhysRevD.81.014029} {\bibfield  {journal} {\bibinfo  {journal} {Phys.
  Rev.}\ }\textbf {\bibinfo {volume} {D81}},\ \bibinfo {pages} {014029}
  (\bibinfo {year} {2010})},\ \Eprint {http://arxiv.org/abs/0911.4407}
  {arXiv:0911.4407 [hep-ph]} \BibitemShut {NoStop}%
\bibitem [{\citenamefont {Hanhart}\ \emph {et~al.}(2012)\citenamefont
  {Hanhart}, \citenamefont {Kalashnikova}, \citenamefont {Kudryavtsev},\ and\
  \citenamefont {Nefediev}}]{Hanhart:2011tn}%
  \BibitemOpen
  \bibfield  {author} {\bibinfo {author} {\bibfnamefont {C.}~\bibnamefont
  {Hanhart}}, \bibinfo {author} {\bibfnamefont {{\relax Yu}.~S.}\ \bibnamefont
  {Kalashnikova}}, \bibinfo {author} {\bibfnamefont {A.~E.}\ \bibnamefont
  {Kudryavtsev}}, \ and\ \bibinfo {author} {\bibfnamefont {A.~V.}\ \bibnamefont
  {Nefediev}},\ }\href {\doibase 10.1103/PhysRevD.85.011501} {\bibfield
  {journal} {\bibinfo  {journal} {Phys. Rev.}\ }\textbf {\bibinfo {volume}
  {D85}},\ \bibinfo {pages} {011501} (\bibinfo {year} {2012})},\ \Eprint
  {http://arxiv.org/abs/1111.6241} {arXiv:1111.6241 [hep-ph]} \BibitemShut
  {NoStop}%
\bibitem [{\citenamefont {Esposito}\ \emph {et~al.}(2021)\citenamefont
  {Esposito}, \citenamefont {Ferreiro}, \citenamefont {Pilloni}, \citenamefont
  {Polosa},\ and\ \citenamefont {Salgado}}]{Esposito:2020ywk}%
  \BibitemOpen
  \bibfield  {author} {\bibinfo {author} {\bibfnamefont {A.}~\bibnamefont
  {Esposito}}, \bibinfo {author} {\bibfnamefont {E.~G.}\ \bibnamefont
  {Ferreiro}}, \bibinfo {author} {\bibfnamefont {A.}~\bibnamefont {Pilloni}},
  \bibinfo {author} {\bibfnamefont {A.~D.}\ \bibnamefont {Polosa}}, \ and\
  \bibinfo {author} {\bibfnamefont {C.~A.}\ \bibnamefont {Salgado}},\ }\href
  {\doibase 10.1140/epjc/s10052-021-09425-w} {\bibfield  {journal} {\bibinfo
  {journal} {Eur. Phys. J. C}\ }\textbf {\bibinfo {volume} {81}},\ \bibinfo
  {pages} {669} (\bibinfo {year} {2021})},\ \Eprint
  {http://arxiv.org/abs/2006.15044} {arXiv:2006.15044 [hep-ph]} \BibitemShut
  {NoStop}%
\bibitem [{\citenamefont {Braaten}\ \emph {et~al.}(2021)\citenamefont
  {Braaten}, \citenamefont {He}, \citenamefont {Ingles},\ and\ \citenamefont
  {Jiang}}]{Braaten:2020iqw}%
  \BibitemOpen
  \bibfield  {author} {\bibinfo {author} {\bibfnamefont {E.}~\bibnamefont
  {Braaten}}, \bibinfo {author} {\bibfnamefont {L.-P.}\ \bibnamefont {He}},
  \bibinfo {author} {\bibfnamefont {K.}~\bibnamefont {Ingles}}, \ and\ \bibinfo
  {author} {\bibfnamefont {J.}~\bibnamefont {Jiang}},\ }\href {\doibase
  10.1103/PhysRevD.103.L071901} {\bibfield  {journal} {\bibinfo  {journal}
  {Phys. Rev. D}\ }\textbf {\bibinfo {volume} {103}},\ \bibinfo {pages}
  {L071901} (\bibinfo {year} {2021})},\ \Eprint
  {http://arxiv.org/abs/2012.13499} {arXiv:2012.13499 [hep-ph]} \BibitemShut
  {NoStop}%
\bibitem [{\citenamefont {Guo}\ \emph {et~al.}(2023)\citenamefont {Guo},
  \citenamefont {Li}, \citenamefont {Zhao},\ and\ \citenamefont
  {He}}]{Guo:2022crh}%
  \BibitemOpen
  \bibfield  {author} {\bibinfo {author} {\bibfnamefont {T.}~\bibnamefont
  {Guo}}, \bibinfo {author} {\bibfnamefont {J.}~\bibnamefont {Li}}, \bibinfo
  {author} {\bibfnamefont {J.}~\bibnamefont {Zhao}}, \ and\ \bibinfo {author}
  {\bibfnamefont {L.}~\bibnamefont {He}},\ }\href {\doibase
  10.1088/1674-1137/accb87} {\bibfield  {journal} {\bibinfo  {journal} {Chin.
  Phys. C}\ }\textbf {\bibinfo {volume} {47}},\ \bibinfo {pages} {063107}
  (\bibinfo {year} {2023})},\ \Eprint {http://arxiv.org/abs/2211.10834}
  {arXiv:2211.10834 [hep-ph]} \BibitemShut {NoStop}%
\bibitem [{\citenamefont {Agaev}\ \emph {et~al.}(2023)\citenamefont {Agaev},
  \citenamefont {Azizi},\ and\ \citenamefont {Sundu}}]{Agaev:2022pis}%
  \BibitemOpen
  \bibfield  {author} {\bibinfo {author} {\bibfnamefont {S.~S.}\ \bibnamefont
  {Agaev}}, \bibinfo {author} {\bibfnamefont {K.}~\bibnamefont {Azizi}}, \ and\
  \bibinfo {author} {\bibfnamefont {H.}~\bibnamefont {Sundu}},\ }\href
  {\doibase 10.1103/PhysRevD.107.054017} {\bibfield  {journal} {\bibinfo
  {journal} {Phys. Rev. D}\ }\textbf {\bibinfo {volume} {107}},\ \bibinfo
  {pages} {054017} (\bibinfo {year} {2023})},\ \Eprint
  {http://arxiv.org/abs/2211.14129} {arXiv:2211.14129 [hep-ph]} \BibitemShut
  {NoStop}%
\bibitem [{\citenamefont {Aaij}\ \emph
  {et~al.}(2022{\natexlab{a}})\citenamefont {Aaij} \emph
  {et~al.}}]{LHCb:2021vvq}%
  \BibitemOpen
  \bibfield  {author} {\bibinfo {author} {\bibfnamefont {R.}~\bibnamefont
  {Aaij}} \emph {et~al.} (\bibinfo {collaboration} {LHCb}),\ }\href {\doibase
  10.1038/s41567-022-01614-y} {\bibfield  {journal} {\bibinfo  {journal}
  {Nature Phys.}\ }\textbf {\bibinfo {volume} {18}},\ \bibinfo {pages} {751}
  (\bibinfo {year} {2022}{\natexlab{a}})},\ \Eprint
  {http://arxiv.org/abs/2109.01038} {arXiv:2109.01038 [hep-ex]} \BibitemShut
  {NoStop}%
\bibitem [{\citenamefont {Aaij}\ \emph
  {et~al.}(2022{\natexlab{b}})\citenamefont {Aaij} \emph
  {et~al.}}]{LHCb:2021auc}%
  \BibitemOpen
  \bibfield  {author} {\bibinfo {author} {\bibfnamefont {R.}~\bibnamefont
  {Aaij}} \emph {et~al.} (\bibinfo {collaboration} {LHCb}),\ }\href {\doibase
  10.1038/s41467-022-30206-w} {\bibfield  {journal} {\bibinfo  {journal}
  {Nature Commun.}\ }\textbf {\bibinfo {volume} {13}},\ \bibinfo {pages} {3351}
  (\bibinfo {year} {2022}{\natexlab{b}})},\ \Eprint
  {http://arxiv.org/abs/2109.01056} {arXiv:2109.01056 [hep-ex]} \BibitemShut
  {NoStop}%
\bibitem [{\citenamefont {Meng}\ \emph {et~al.}(2021)\citenamefont {Meng},
  \citenamefont {Wang}, \citenamefont {Wang},\ and\ \citenamefont
  {Zhu}}]{Meng:2021jnw}%
  \BibitemOpen
  \bibfield  {author} {\bibinfo {author} {\bibfnamefont {L.}~\bibnamefont
  {Meng}}, \bibinfo {author} {\bibfnamefont {G.-J.}\ \bibnamefont {Wang}},
  \bibinfo {author} {\bibfnamefont {B.}~\bibnamefont {Wang}}, \ and\ \bibinfo
  {author} {\bibfnamefont {S.-L.}\ \bibnamefont {Zhu}},\ }\href {\doibase
  10.1103/PhysRevD.104.L051502} {\bibfield  {journal} {\bibinfo  {journal}
  {Phys. Rev. D}\ }\textbf {\bibinfo {volume} {104}},\ \bibinfo {pages}
  {051502} (\bibinfo {year} {2021})},\ \Eprint
  {http://arxiv.org/abs/2107.14784} {arXiv:2107.14784 [hep-ph]} \BibitemShut
  {NoStop}%
\bibitem [{\citenamefont {Agaev}\ \emph
  {et~al.}(2022{\natexlab{a}})\citenamefont {Agaev}, \citenamefont {Azizi},\
  and\ \citenamefont {Sundu}}]{Agaev:2021vur}%
  \BibitemOpen
  \bibfield  {author} {\bibinfo {author} {\bibfnamefont {S.~S.}\ \bibnamefont
  {Agaev}}, \bibinfo {author} {\bibfnamefont {K.}~\bibnamefont {Azizi}}, \ and\
  \bibinfo {author} {\bibfnamefont {H.}~\bibnamefont {Sundu}},\ }\href
  {\doibase 10.1016/j.nuclphysb.2022.115650} {\bibfield  {journal} {\bibinfo
  {journal} {Nucl. Phys. B}\ }\textbf {\bibinfo {volume} {975}},\ \bibinfo
  {pages} {115650} (\bibinfo {year} {2022}{\natexlab{a}})},\ \Eprint
  {http://arxiv.org/abs/2108.00188} {arXiv:2108.00188 [hep-ph]} \BibitemShut
  {NoStop}%
\bibitem [{\citenamefont {Ling}\ \emph {et~al.}(2022)\citenamefont {Ling},
  \citenamefont {Liu}, \citenamefont {Geng}, \citenamefont {Wang},\ and\
  \citenamefont {Xie}}]{Ling:2021bir}%
  \BibitemOpen
  \bibfield  {author} {\bibinfo {author} {\bibfnamefont {X.-Z.}\ \bibnamefont
  {Ling}}, \bibinfo {author} {\bibfnamefont {M.-Z.}\ \bibnamefont {Liu}},
  \bibinfo {author} {\bibfnamefont {L.-S.}\ \bibnamefont {Geng}}, \bibinfo
  {author} {\bibfnamefont {E.}~\bibnamefont {Wang}}, \ and\ \bibinfo {author}
  {\bibfnamefont {J.-J.}\ \bibnamefont {Xie}},\ }\href {\doibase
  10.1016/j.physletb.2022.136897} {\bibfield  {journal} {\bibinfo  {journal}
  {Phys. Lett. B}\ }\textbf {\bibinfo {volume} {826}},\ \bibinfo {pages}
  {136897} (\bibinfo {year} {2022})},\ \Eprint
  {http://arxiv.org/abs/2108.00947} {arXiv:2108.00947 [hep-ph]} \BibitemShut
  {NoStop}%
\bibitem [{\citenamefont {Dong}\ \emph
  {et~al.}(2021{\natexlab{a}})\citenamefont {Dong}, \citenamefont {Guo},\ and\
  \citenamefont {Zou}}]{Dong:2021bvy}%
  \BibitemOpen
  \bibfield  {author} {\bibinfo {author} {\bibfnamefont {X.-K.}\ \bibnamefont
  {Dong}}, \bibinfo {author} {\bibfnamefont {F.-K.}\ \bibnamefont {Guo}}, \
  and\ \bibinfo {author} {\bibfnamefont {B.-S.}\ \bibnamefont {Zou}},\ }\href
  {\doibase 10.1088/1572-9494/ac27a2} {\bibfield  {journal} {\bibinfo
  {journal} {Commun. Theor. Phys.}\ }\textbf {\bibinfo {volume} {73}},\
  \bibinfo {pages} {125201} (\bibinfo {year} {2021}{\natexlab{a}})},\ \Eprint
  {http://arxiv.org/abs/2108.02673} {arXiv:2108.02673 [hep-ph]} \BibitemShut
  {NoStop}%
\bibitem [{\citenamefont {Feijoo}\ \emph {et~al.}(2021)\citenamefont {Feijoo},
  \citenamefont {Liang},\ and\ \citenamefont {Oset}}]{Feijoo:2021ppq}%
  \BibitemOpen
  \bibfield  {author} {\bibinfo {author} {\bibfnamefont {A.}~\bibnamefont
  {Feijoo}}, \bibinfo {author} {\bibfnamefont {W.~H.}\ \bibnamefont {Liang}}, \
  and\ \bibinfo {author} {\bibfnamefont {E.}~\bibnamefont {Oset}},\ }\href
  {\doibase 10.1103/PhysRevD.104.114015} {\bibfield  {journal} {\bibinfo
  {journal} {Phys. Rev. D}\ }\textbf {\bibinfo {volume} {104}},\ \bibinfo
  {pages} {114015} (\bibinfo {year} {2021})},\ \Eprint
  {http://arxiv.org/abs/2108.02730} {arXiv:2108.02730 [hep-ph]} \BibitemShut
  {NoStop}%
\bibitem [{\citenamefont {Fleming}\ \emph {et~al.}(2021)\citenamefont
  {Fleming}, \citenamefont {Hodges},\ and\ \citenamefont
  {Mehen}}]{Fleming:2021wmk}%
  \BibitemOpen
  \bibfield  {author} {\bibinfo {author} {\bibfnamefont {S.}~\bibnamefont
  {Fleming}}, \bibinfo {author} {\bibfnamefont {R.}~\bibnamefont {Hodges}}, \
  and\ \bibinfo {author} {\bibfnamefont {T.}~\bibnamefont {Mehen}},\ }\href
  {\doibase 10.1103/PhysRevD.104.116010} {\bibfield  {journal} {\bibinfo
  {journal} {Phys. Rev. D}\ }\textbf {\bibinfo {volume} {104}},\ \bibinfo
  {pages} {116010} (\bibinfo {year} {2021})},\ \Eprint
  {http://arxiv.org/abs/2109.02188} {arXiv:2109.02188 [hep-ph]} \BibitemShut
  {NoStop}%
\bibitem [{\citenamefont {Agaev}\ \emph
  {et~al.}(2022{\natexlab{b}})\citenamefont {Agaev}, \citenamefont {Azizi},\
  and\ \citenamefont {Sundu}}]{Agaev:2022ast}%
  \BibitemOpen
  \bibfield  {author} {\bibinfo {author} {\bibfnamefont {S.~S.}\ \bibnamefont
  {Agaev}}, \bibinfo {author} {\bibfnamefont {K.}~\bibnamefont {Azizi}}, \ and\
  \bibinfo {author} {\bibfnamefont {H.}~\bibnamefont {Sundu}},\ }\href
  {\doibase 10.1007/JHEP06(2022)057} {\bibfield  {journal} {\bibinfo  {journal}
  {JHEP}\ }\textbf {\bibinfo {volume} {06}},\ \bibinfo {pages} {057} (\bibinfo
  {year} {2022}{\natexlab{b}})},\ \Eprint {http://arxiv.org/abs/2201.02788}
  {arXiv:2201.02788 [hep-ph]} \BibitemShut {NoStop}%
\bibitem [{\citenamefont {Carlson}\ \emph {et~al.}(1988)\citenamefont
  {Carlson}, \citenamefont {Heller},\ and\ \citenamefont
  {Tjon}}]{Carlson:1987hh}%
  \BibitemOpen
  \bibfield  {author} {\bibinfo {author} {\bibfnamefont {J.}~\bibnamefont
  {Carlson}}, \bibinfo {author} {\bibfnamefont {L.}~\bibnamefont {Heller}}, \
  and\ \bibinfo {author} {\bibfnamefont {J.~A.}\ \bibnamefont {Tjon}},\ }\href
  {\doibase 10.1103/PhysRevD.37.744} {\bibfield  {journal} {\bibinfo  {journal}
  {Phys. Rev.}\ }\textbf {\bibinfo {volume} {D37}},\ \bibinfo {pages} {744}
  (\bibinfo {year} {1988})}\BibitemShut {NoStop}%
\bibitem [{\citenamefont {Silvestre-Brac}\ and\ \citenamefont
  {Semay}(1993)}]{Silvestre-Brac:1993zem}%
  \BibitemOpen
  \bibfield  {author} {\bibinfo {author} {\bibfnamefont {B.}~\bibnamefont
  {Silvestre-Brac}}\ and\ \bibinfo {author} {\bibfnamefont {C.}~\bibnamefont
  {Semay}},\ }\href {\doibase 10.1007/BF01565058} {\bibfield  {journal}
  {\bibinfo  {journal} {Z. Phys. C}\ }\textbf {\bibinfo {volume} {57}},\
  \bibinfo {pages} {273} (\bibinfo {year} {1993})}\BibitemShut {NoStop}%
\bibitem [{\citenamefont {Semay}\ and\ \citenamefont
  {Silvestre-Brac}(1994)}]{Semay:1994ht}%
  \BibitemOpen
  \bibfield  {author} {\bibinfo {author} {\bibfnamefont {C.}~\bibnamefont
  {Semay}}\ and\ \bibinfo {author} {\bibfnamefont {B.}~\bibnamefont
  {Silvestre-Brac}},\ }\href {\doibase 10.1007/BF01413104} {\bibfield
  {journal} {\bibinfo  {journal} {Z. Phys. C}\ }\textbf {\bibinfo {volume}
  {61}},\ \bibinfo {pages} {271} (\bibinfo {year} {1994})}\BibitemShut
  {NoStop}%
\bibitem [{\citenamefont {Pepin}\ \emph {et~al.}(1997)\citenamefont {Pepin},
  \citenamefont {Stancu}, \citenamefont {Genovese},\ and\ \citenamefont
  {Richard}}]{Pepin:1996id}%
  \BibitemOpen
  \bibfield  {author} {\bibinfo {author} {\bibfnamefont {S.}~\bibnamefont
  {Pepin}}, \bibinfo {author} {\bibfnamefont {F.}~\bibnamefont {Stancu}},
  \bibinfo {author} {\bibfnamefont {M.}~\bibnamefont {Genovese}}, \ and\
  \bibinfo {author} {\bibfnamefont {J.~M.}\ \bibnamefont {Richard}},\ }\href
  {\doibase 10.1016/S0370-2693(96)01597-3} {\bibfield  {journal} {\bibinfo
  {journal} {Phys. Lett. B}\ }\textbf {\bibinfo {volume} {393}},\ \bibinfo
  {pages} {119} (\bibinfo {year} {1997})},\ \Eprint
  {http://arxiv.org/abs/hep-ph/9609348} {arXiv:hep-ph/9609348} \BibitemShut
  {NoStop}%
\bibitem [{\citenamefont {Peng}\ \emph {et~al.}(2022)\citenamefont {Peng},
  \citenamefont {S\'anchez~S\'anchez}, \citenamefont {Yan},\ and\ \citenamefont
  {Pavon~Valderrama}}]{Peng:2021hkr}%
  \BibitemOpen
  \bibfield  {author} {\bibinfo {author} {\bibfnamefont {F.-Z.}\ \bibnamefont
  {Peng}}, \bibinfo {author} {\bibfnamefont {M.}~\bibnamefont
  {S\'anchez~S\'anchez}}, \bibinfo {author} {\bibfnamefont {M.-J.}\
  \bibnamefont {Yan}}, \ and\ \bibinfo {author} {\bibfnamefont
  {M.}~\bibnamefont {Pavon~Valderrama}},\ }\href {\doibase
  10.1103/PhysRevD.105.034028} {\bibfield  {journal} {\bibinfo  {journal}
  {Phys. Rev. D}\ }\textbf {\bibinfo {volume} {105}},\ \bibinfo {pages}
  {034028} (\bibinfo {year} {2022})},\ \Eprint
  {http://arxiv.org/abs/2101.07213} {arXiv:2101.07213 [hep-ph]} \BibitemShut
  {NoStop}%
\bibitem [{\citenamefont {Janc}\ and\ \citenamefont
  {Rosina}(2004)}]{Janc:2004qn}%
  \BibitemOpen
  \bibfield  {author} {\bibinfo {author} {\bibfnamefont {D.}~\bibnamefont
  {Janc}}\ and\ \bibinfo {author} {\bibfnamefont {M.}~\bibnamefont {Rosina}},\
  }\href {\doibase 10.1007/s00601-004-0068-9} {\bibfield  {journal} {\bibinfo
  {journal} {Few Body Syst.}\ }\textbf {\bibinfo {volume} {35}},\ \bibinfo
  {pages} {175} (\bibinfo {year} {2004})},\ \Eprint
  {http://arxiv.org/abs/hep-ph/0405208} {arXiv:hep-ph/0405208} \BibitemShut
  {NoStop}%
\bibitem [{\citenamefont {Valderrama}(2012)}]{Valderrama:2012jv}%
  \BibitemOpen
  \bibfield  {author} {\bibinfo {author} {\bibfnamefont {M.~P.}\ \bibnamefont
  {Valderrama}},\ }\href {\doibase 10.1103/PhysRevD.85.114037} {\bibfield
  {journal} {\bibinfo  {journal} {Phys. Rev.}\ }\textbf {\bibinfo {volume}
  {D85}},\ \bibinfo {pages} {114037} (\bibinfo {year} {2012})},\ \Eprint
  {http://arxiv.org/abs/1204.2400} {arXiv:1204.2400 [hep-ph]} \BibitemShut
  {NoStop}%
\bibitem [{\citenamefont {Nieves}\ and\ \citenamefont
  {Valderrama}(2012)}]{Nieves:2012tt}%
  \BibitemOpen
  \bibfield  {author} {\bibinfo {author} {\bibfnamefont {J.}~\bibnamefont
  {Nieves}}\ and\ \bibinfo {author} {\bibfnamefont {M.~P.}\ \bibnamefont
  {Valderrama}},\ }\href {\doibase 10.1103/PhysRevD.86.056004} {\bibfield
  {journal} {\bibinfo  {journal} {Phys. Rev.}\ }\textbf {\bibinfo {volume}
  {D86}},\ \bibinfo {pages} {056004} (\bibinfo {year} {2012})},\ \Eprint
  {http://arxiv.org/abs/1204.2790} {arXiv:1204.2790 [hep-ph]} \BibitemShut
  {NoStop}%
\bibitem [{\citenamefont {Peng}\ \emph {et~al.}(2020)\citenamefont {Peng},
  \citenamefont {Liu}, \citenamefont {S\'anchez~S\'anchez},\ and\ \citenamefont
  {Pavon~Valderrama}}]{Peng:2020xrf}%
  \BibitemOpen
  \bibfield  {author} {\bibinfo {author} {\bibfnamefont {F.-Z.}\ \bibnamefont
  {Peng}}, \bibinfo {author} {\bibfnamefont {M.-Z.}\ \bibnamefont {Liu}},
  \bibinfo {author} {\bibfnamefont {M.}~\bibnamefont {S\'anchez~S\'anchez}}, \
  and\ \bibinfo {author} {\bibfnamefont {M.}~\bibnamefont {Pavon~Valderrama}},\
  }\href {\doibase 10.1103/PhysRevD.102.114020} {\bibfield  {journal} {\bibinfo
   {journal} {Phys. Rev. D}\ }\textbf {\bibinfo {volume} {102}},\ \bibinfo
  {pages} {114020} (\bibinfo {year} {2020})},\ \Eprint
  {http://arxiv.org/abs/2004.05658} {arXiv:2004.05658 [hep-ph]} \BibitemShut
  {NoStop}%
\bibitem [{\citenamefont {Pav\'on~Valderrama}\ and\ \citenamefont
  {Phillips}(2015)}]{PavonValderrama:2014zeq}%
  \BibitemOpen
  \bibfield  {author} {\bibinfo {author} {\bibfnamefont {M.}~\bibnamefont
  {Pav\'on~Valderrama}}\ and\ \bibinfo {author} {\bibfnamefont {D.~R.}\
  \bibnamefont {Phillips}},\ }\href {\doibase 10.1103/PhysRevLett.114.082502}
  {\bibfield  {journal} {\bibinfo  {journal} {Phys. Rev. Lett.}\ }\textbf
  {\bibinfo {volume} {114}},\ \bibinfo {pages} {082502} (\bibinfo {year}
  {2015})},\ \Eprint {http://arxiv.org/abs/1407.0437} {arXiv:1407.0437
  [nucl-th]} \BibitemShut {NoStop}%
\bibitem [{\citenamefont {Langer}(1937)}]{Langer:1937qr}%
  \BibitemOpen
  \bibfield  {author} {\bibinfo {author} {\bibfnamefont {R.~E.}\ \bibnamefont
  {Langer}},\ }\href {\doibase 10.1103/PhysRev.51.669} {\bibfield  {journal}
  {\bibinfo  {journal} {Phys. Rev.}\ }\textbf {\bibinfo {volume} {51}},\
  \bibinfo {pages} {669} (\bibinfo {year} {1937})}\BibitemShut {NoStop}%
\bibitem [{\citenamefont {Prelovsek}\ \emph {et~al.}(2021)\citenamefont
  {Prelovsek}, \citenamefont {Collins}, \citenamefont {Mohler}, \citenamefont
  {Padmanath},\ and\ \citenamefont {Piemonte}}]{Prelovsek:2020eiw}%
  \BibitemOpen
  \bibfield  {author} {\bibinfo {author} {\bibfnamefont {S.}~\bibnamefont
  {Prelovsek}}, \bibinfo {author} {\bibfnamefont {S.}~\bibnamefont {Collins}},
  \bibinfo {author} {\bibfnamefont {D.}~\bibnamefont {Mohler}}, \bibinfo
  {author} {\bibfnamefont {M.}~\bibnamefont {Padmanath}}, \ and\ \bibinfo
  {author} {\bibfnamefont {S.}~\bibnamefont {Piemonte}},\ }\href {\doibase
  10.1007/JHEP06(2021)035} {\bibfield  {journal} {\bibinfo  {journal} {JHEP}\
  }\textbf {\bibinfo {volume} {06}},\ \bibinfo {pages} {035} (\bibinfo {year}
  {2021})},\ \Eprint {http://arxiv.org/abs/2011.02542} {arXiv:2011.02542
  [hep-lat]} \BibitemShut {NoStop}%
\bibitem [{\citenamefont {Sakurai}(1960)}]{Sakurai:1960ju}%
  \BibitemOpen
  \bibfield  {author} {\bibinfo {author} {\bibfnamefont {J.~J.}\ \bibnamefont
  {Sakurai}},\ }\href {\doibase 10.1016/0003-4916(60)90126-3} {\bibfield
  {journal} {\bibinfo  {journal} {Annals Phys.}\ }\textbf {\bibinfo {volume}
  {11}},\ \bibinfo {pages} {1} (\bibinfo {year} {1960})}\BibitemShut {NoStop}%
\bibitem [{\citenamefont {Kawarabayashi}\ and\ \citenamefont
  {Suzuki}(1966)}]{Kawarabayashi:1966kd}%
  \BibitemOpen
  \bibfield  {author} {\bibinfo {author} {\bibfnamefont {K.}~\bibnamefont
  {Kawarabayashi}}\ and\ \bibinfo {author} {\bibfnamefont {M.}~\bibnamefont
  {Suzuki}},\ }\href {\doibase 10.1103/PhysRevLett.16.255} {\bibfield
  {journal} {\bibinfo  {journal} {Phys. Rev. Lett.}\ }\textbf {\bibinfo
  {volume} {16}},\ \bibinfo {pages} {255} (\bibinfo {year} {1966})}\BibitemShut
  {NoStop}%
\bibitem [{\citenamefont {Riazuddin}\ and\ \citenamefont
  {Fayyazuddin}(1966)}]{Riazuddin:1966sw}%
  \BibitemOpen
  \bibfield  {author} {\bibinfo {author} {\bibnamefont {Riazuddin}}\ and\
  \bibinfo {author} {\bibnamefont {Fayyazuddin}},\ }\href {\doibase
  10.1103/PhysRev.147.1071} {\bibfield  {journal} {\bibinfo  {journal} {Phys.
  Rev.}\ }\textbf {\bibinfo {volume} {147}},\ \bibinfo {pages} {1071} (\bibinfo
  {year} {1966})}\BibitemShut {NoStop}%
\bibitem [{\citenamefont {Gell-Mann}\ and\ \citenamefont
  {Levy}(1960)}]{GellMann:1960np}%
  \BibitemOpen
  \bibfield  {author} {\bibinfo {author} {\bibfnamefont {M.}~\bibnamefont
  {Gell-Mann}}\ and\ \bibinfo {author} {\bibfnamefont {M.}~\bibnamefont
  {Levy}},\ }\href {\doibase 10.1007/BF02859738} {\bibfield  {journal}
  {\bibinfo  {journal} {Nuovo Cim.}\ }\textbf {\bibinfo {volume} {16}},\
  \bibinfo {pages} {705} (\bibinfo {year} {1960})}\BibitemShut {NoStop}%
\bibitem [{\citenamefont {Yan}\ \emph {et~al.}(2021)\citenamefont {Yan},
  \citenamefont {Peng}, \citenamefont {S\'anchez~S\'anchez},\ and\
  \citenamefont {Pavon~Valderrama}}]{Yan:2021tcp}%
  \BibitemOpen
  \bibfield  {author} {\bibinfo {author} {\bibfnamefont {M.-J.}\ \bibnamefont
  {Yan}}, \bibinfo {author} {\bibfnamefont {F.-Z.}\ \bibnamefont {Peng}},
  \bibinfo {author} {\bibfnamefont {M.}~\bibnamefont {S\'anchez~S\'anchez}}, \
  and\ \bibinfo {author} {\bibfnamefont {M.}~\bibnamefont {Pavon~Valderrama}},\
  }\href {\doibase 10.1103/PhysRevD.104.114025} {\bibfield  {journal} {\bibinfo
   {journal} {Phys. Rev. D}\ }\textbf {\bibinfo {volume} {104}},\ \bibinfo
  {pages} {114025} (\bibinfo {year} {2021})},\ \Eprint
  {http://arxiv.org/abs/2102.13058} {arXiv:2102.13058 [hep-ph]} \BibitemShut
  {NoStop}%
\bibitem [{\citenamefont {Machleidt}\ \emph {et~al.}(1987)\citenamefont
  {Machleidt}, \citenamefont {Holinde},\ and\ \citenamefont
  {Elster}}]{Machleidt:1987hj}%
  \BibitemOpen
  \bibfield  {author} {\bibinfo {author} {\bibfnamefont {R.}~\bibnamefont
  {Machleidt}}, \bibinfo {author} {\bibfnamefont {K.}~\bibnamefont {Holinde}},
  \ and\ \bibinfo {author} {\bibfnamefont {C.}~\bibnamefont {Elster}},\ }\href
  {\doibase 10.1016/S0370-1573(87)80002-9} {\bibfield  {journal} {\bibinfo
  {journal} {Phys. Rept.}\ }\textbf {\bibinfo {volume} {149}},\ \bibinfo
  {pages} {1} (\bibinfo {year} {1987})}\BibitemShut {NoStop}%
\bibitem [{\citenamefont {Machleidt}(1989)}]{Machleidt:1989tm}%
  \BibitemOpen
  \bibfield  {author} {\bibinfo {author} {\bibfnamefont {R.}~\bibnamefont
  {Machleidt}},\ }\href@noop {} {\bibfield  {journal} {\bibinfo  {journal}
  {Adv. Nucl. Phys.}\ }\textbf {\bibinfo {volume} {19}},\ \bibinfo {pages}
  {189} (\bibinfo {year} {1989})}\BibitemShut {NoStop}%
\bibitem [{\citenamefont {Binstock}\ and\ \citenamefont
  {Bryan}(1971)}]{Binstock:1971duy}%
  \BibitemOpen
  \bibfield  {author} {\bibinfo {author} {\bibfnamefont {J.}~\bibnamefont
  {Binstock}}\ and\ \bibinfo {author} {\bibfnamefont {R.}~\bibnamefont
  {Bryan}},\ }\href {\doibase 10.1103/PhysRevD.4.1341} {\bibfield  {journal}
  {\bibinfo  {journal} {Phys. Rev. D}\ }\textbf {\bibinfo {volume} {4}},\
  \bibinfo {pages} {1341} (\bibinfo {year} {1971})}\BibitemShut {NoStop}%
\bibitem [{\citenamefont {Flambaum}\ and\ \citenamefont
  {Shuryak}(2007)}]{Flambaum:2007xj}%
  \BibitemOpen
  \bibfield  {author} {\bibinfo {author} {\bibfnamefont {V.~V.}\ \bibnamefont
  {Flambaum}}\ and\ \bibinfo {author} {\bibfnamefont {E.~V.}\ \bibnamefont
  {Shuryak}},\ }\href {\doibase 10.1103/PhysRevC.76.065206} {\bibfield
  {journal} {\bibinfo  {journal} {Phys. Rev. C}\ }\textbf {\bibinfo {volume}
  {76}},\ \bibinfo {pages} {065206} (\bibinfo {year} {2007})},\ \Eprint
  {http://arxiv.org/abs/nucl-th/0702038} {arXiv:nucl-th/0702038} \BibitemShut
  {NoStop}%
\bibitem [{\citenamefont {Liu}\ \emph {et~al.}(2019)\citenamefont {Liu},
  \citenamefont {Wu}, \citenamefont {Pavon~Valderrama}, \citenamefont {Xie},\
  and\ \citenamefont {Geng}}]{Liu:2019stu}%
  \BibitemOpen
  \bibfield  {author} {\bibinfo {author} {\bibfnamefont {M.-Z.}\ \bibnamefont
  {Liu}}, \bibinfo {author} {\bibfnamefont {T.-W.}\ \bibnamefont {Wu}},
  \bibinfo {author} {\bibfnamefont {M.}~\bibnamefont {Pavon~Valderrama}},
  \bibinfo {author} {\bibfnamefont {J.-J.}\ \bibnamefont {Xie}}, \ and\
  \bibinfo {author} {\bibfnamefont {L.-S.}\ \bibnamefont {Geng}},\ }\href
  {\doibase 10.1103/PhysRevD.99.094018} {\bibfield  {journal} {\bibinfo
  {journal} {Phys. Rev.}\ }\textbf {\bibinfo {volume} {D99}},\ \bibinfo {pages}
  {094018} (\bibinfo {year} {2019})},\ \Eprint
  {http://arxiv.org/abs/1902.03044} {arXiv:1902.03044 [hep-ph]} \BibitemShut
  {NoStop}%
\bibitem [{\citenamefont {Hanhart}\ \emph {et~al.}(2015)\citenamefont
  {Hanhart}, \citenamefont {Kalashnikova}, \citenamefont {Matuschek},
  \citenamefont {Mizuk}, \citenamefont {Nefediev},\ and\ \citenamefont
  {Wang}}]{Hanhart:2015cua}%
  \BibitemOpen
  \bibfield  {author} {\bibinfo {author} {\bibfnamefont {C.}~\bibnamefont
  {Hanhart}}, \bibinfo {author} {\bibfnamefont {Y.~S.}\ \bibnamefont
  {Kalashnikova}}, \bibinfo {author} {\bibfnamefont {P.}~\bibnamefont
  {Matuschek}}, \bibinfo {author} {\bibfnamefont {R.~V.}\ \bibnamefont
  {Mizuk}}, \bibinfo {author} {\bibfnamefont {A.~V.}\ \bibnamefont {Nefediev}},
  \ and\ \bibinfo {author} {\bibfnamefont {Q.}~\bibnamefont {Wang}},\ }\href
  {\doibase 10.1103/PhysRevLett.115.202001} {\bibfield  {journal} {\bibinfo
  {journal} {Phys. Rev. Lett.}\ }\textbf {\bibinfo {volume} {115}},\ \bibinfo
  {pages} {202001} (\bibinfo {year} {2015})},\ \Eprint
  {http://arxiv.org/abs/1507.00382} {arXiv:1507.00382 [hep-ph]} \BibitemShut
  {NoStop}%
\bibitem [{\citenamefont {Dong}\ \emph
  {et~al.}(2021{\natexlab{b}})\citenamefont {Dong}, \citenamefont {Guo},\ and\
  \citenamefont {Zou}}]{Dong:2020hxe}%
  \BibitemOpen
  \bibfield  {author} {\bibinfo {author} {\bibfnamefont {X.-K.}\ \bibnamefont
  {Dong}}, \bibinfo {author} {\bibfnamefont {F.-K.}\ \bibnamefont {Guo}}, \
  and\ \bibinfo {author} {\bibfnamefont {B.-S.}\ \bibnamefont {Zou}},\ }\href
  {\doibase 10.1103/PhysRevLett.126.152001} {\bibfield  {journal} {\bibinfo
  {journal} {Phys. Rev. Lett.}\ }\textbf {\bibinfo {volume} {126}},\ \bibinfo
  {pages} {152001} (\bibinfo {year} {2021}{\natexlab{b}})},\ \Eprint
  {http://arxiv.org/abs/2011.14517} {arXiv:2011.14517 [hep-ph]} \BibitemShut
  {NoStop}%
\bibitem [{\citenamefont {Chen}\ and\ \citenamefont
  {Sparado~Norella}(2022)}]{lhcb2022-a}%
  \BibitemOpen
  \bibfield  {author} {\bibinfo {author} {\bibfnamefont {C.}~\bibnamefont
  {Chen}}\ and\ \bibinfo {author} {\bibfnamefont {E.}~\bibnamefont
  {Sparado~Norella}},\ }\href@noop {} {\enquote {\bibinfo {title} {Particle zoo
  2.0: New tetra- and pentaquarks at lhcb},}\ }\bibinfo {howpublished}
  {\url{https://indico.cern.ch/event/1176505/}} (\bibinfo {year} {2022}),\
  \bibinfo {note} {presented at the LHC Seminar}\BibitemShut {NoStop}%
\bibitem [{\citenamefont {Maiani}\ \emph {et~al.}(2019)\citenamefont {Maiani},
  \citenamefont {Polosa},\ and\ \citenamefont {Riquer}}]{Maiani:2019lpu}%
  \BibitemOpen
  \bibfield  {author} {\bibinfo {author} {\bibfnamefont {L.}~\bibnamefont
  {Maiani}}, \bibinfo {author} {\bibfnamefont {A.~D.}\ \bibnamefont {Polosa}},
  \ and\ \bibinfo {author} {\bibfnamefont {V.}~\bibnamefont {Riquer}},\ }\href
  {\doibase 10.1103/PhysRevD.100.074002} {\bibfield  {journal} {\bibinfo
  {journal} {Phys. Rev. D}\ }\textbf {\bibinfo {volume} {100}},\ \bibinfo
  {pages} {074002} (\bibinfo {year} {2019})},\ \Eprint
  {http://arxiv.org/abs/1908.03244} {arXiv:1908.03244 [hep-ph]} \BibitemShut
  {NoStop}%
\bibitem [{\citenamefont {Luo}\ \emph {et~al.}(2017)\citenamefont {Luo},
  \citenamefont {Chen}, \citenamefont {Liu}, \citenamefont {Liu},\ and\
  \citenamefont {Zhu}}]{Luo:2017eub}%
  \BibitemOpen
  \bibfield  {author} {\bibinfo {author} {\bibfnamefont {S.-Q.}\ \bibnamefont
  {Luo}}, \bibinfo {author} {\bibfnamefont {K.}~\bibnamefont {Chen}}, \bibinfo
  {author} {\bibfnamefont {X.}~\bibnamefont {Liu}}, \bibinfo {author}
  {\bibfnamefont {Y.-R.}\ \bibnamefont {Liu}}, \ and\ \bibinfo {author}
  {\bibfnamefont {S.-L.}\ \bibnamefont {Zhu}},\ }\href {\doibase
  10.1140/epjc/s10052-017-5297-4} {\bibfield  {journal} {\bibinfo  {journal}
  {Eur. Phys. J. C}\ }\textbf {\bibinfo {volume} {77}},\ \bibinfo {pages} {709}
  (\bibinfo {year} {2017})},\ \Eprint {http://arxiv.org/abs/1707.01180}
  {arXiv:1707.01180 [hep-ph]} \BibitemShut {NoStop}%
\bibitem [{\citenamefont {Wang}\ \emph {et~al.}(2022)\citenamefont {Wang} \emph
  {et~al.}}]{Belle:2021nuv}%
  \BibitemOpen
  \bibfield  {author} {\bibinfo {author} {\bibfnamefont {X.~L.}\ \bibnamefont
  {Wang}} \emph {et~al.} (\bibinfo {collaboration} {Belle}),\ }\href {\doibase
  10.1103/PhysRevD.105.112011} {\bibfield  {journal} {\bibinfo  {journal}
  {Phys. Rev. D}\ }\textbf {\bibinfo {volume} {105}},\ \bibinfo {pages}
  {112011} (\bibinfo {year} {2022})},\ \Eprint
  {http://arxiv.org/abs/2105.06605} {arXiv:2105.06605 [hep-ex]} \BibitemShut
  {NoStop}%
\bibitem [{\citenamefont {Albaladejo}\ \emph {et~al.}(2015)\citenamefont
  {Albaladejo}, \citenamefont {Guo}, \citenamefont {Hidalgo-Duque},
  \citenamefont {Nieves},\ and\ \citenamefont
  {Valderrama}}]{Albaladejo:2015dsa}%
  \BibitemOpen
  \bibfield  {author} {\bibinfo {author} {\bibfnamefont {M.}~\bibnamefont
  {Albaladejo}}, \bibinfo {author} {\bibfnamefont {F.~K.}\ \bibnamefont {Guo}},
  \bibinfo {author} {\bibfnamefont {C.}~\bibnamefont {Hidalgo-Duque}}, \bibinfo
  {author} {\bibfnamefont {J.}~\bibnamefont {Nieves}}, \ and\ \bibinfo {author}
  {\bibfnamefont {M.~P.}\ \bibnamefont {Valderrama}},\ }\href {\doibase
  10.1140/epjc/s10052-015-3753-6} {\bibfield  {journal} {\bibinfo  {journal}
  {Eur. Phys. J.}\ }\textbf {\bibinfo {volume} {C75}},\ \bibinfo {pages} {547}
  (\bibinfo {year} {2015})},\ \Eprint {http://arxiv.org/abs/1504.00861}
  {arXiv:1504.00861 [hep-ph]} \BibitemShut {NoStop}%
\bibitem [{\citenamefont {Albaladejo}\ \emph {et~al.}(2016)\citenamefont
  {Albaladejo}, \citenamefont {Guo}, \citenamefont {Hidalgo-Duque},\ and\
  \citenamefont {Nieves}}]{Albaladejo:2015lob}%
  \BibitemOpen
  \bibfield  {author} {\bibinfo {author} {\bibfnamefont {M.}~\bibnamefont
  {Albaladejo}}, \bibinfo {author} {\bibfnamefont {F.-K.}\ \bibnamefont {Guo}},
  \bibinfo {author} {\bibfnamefont {C.}~\bibnamefont {Hidalgo-Duque}}, \ and\
  \bibinfo {author} {\bibfnamefont {J.}~\bibnamefont {Nieves}},\ }\href
  {\doibase 10.1016/j.physletb.2016.02.025} {\bibfield  {journal} {\bibinfo
  {journal} {Phys. Lett. B}\ }\textbf {\bibinfo {volume} {755}},\ \bibinfo
  {pages} {337} (\bibinfo {year} {2016})},\ \Eprint
  {http://arxiv.org/abs/1512.03638} {arXiv:1512.03638 [hep-ph]} \BibitemShut
  {NoStop}%
\bibitem [{\citenamefont {Yang}\ \emph {et~al.}(2021)\citenamefont {Yang},
  \citenamefont {Cao}, \citenamefont {Guo}, \citenamefont {Nieves},\ and\
  \citenamefont {Valderrama}}]{Yang:2020nrt}%
  \BibitemOpen
  \bibfield  {author} {\bibinfo {author} {\bibfnamefont {Z.}~\bibnamefont
  {Yang}}, \bibinfo {author} {\bibfnamefont {X.}~\bibnamefont {Cao}}, \bibinfo
  {author} {\bibfnamefont {F.-K.}\ \bibnamefont {Guo}}, \bibinfo {author}
  {\bibfnamefont {J.}~\bibnamefont {Nieves}}, \ and\ \bibinfo {author}
  {\bibfnamefont {M.~P.}\ \bibnamefont {Valderrama}},\ }\href {\doibase
  10.1103/PhysRevD.103.074029} {\bibfield  {journal} {\bibinfo  {journal}
  {Phys. Rev. D}\ }\textbf {\bibinfo {volume} {103}},\ \bibinfo {pages}
  {074029} (\bibinfo {year} {2021})},\ \Eprint
  {http://arxiv.org/abs/2011.08725} {arXiv:2011.08725 [hep-ph]} \BibitemShut
  {NoStop}%
\bibitem [{\citenamefont {Godfrey}\ and\ \citenamefont
  {Isgur}(1985)}]{Godfrey:1985xj}%
  \BibitemOpen
  \bibfield  {author} {\bibinfo {author} {\bibfnamefont {S.}~\bibnamefont
  {Godfrey}}\ and\ \bibinfo {author} {\bibfnamefont {N.}~\bibnamefont
  {Isgur}},\ }\href {\doibase 10.1103/PhysRevD.32.189} {\bibfield  {journal}
  {\bibinfo  {journal} {Phys. Rev. D}\ }\textbf {\bibinfo {volume} {32}},\
  \bibinfo {pages} {189} (\bibinfo {year} {1985})}\BibitemShut {NoStop}%
\bibitem [{\citenamefont {Zeng}\ \emph {et~al.}(1995)\citenamefont {Zeng},
  \citenamefont {Van~Orden},\ and\ \citenamefont {Roberts}}]{Zeng:1994vj}%
  \BibitemOpen
  \bibfield  {author} {\bibinfo {author} {\bibfnamefont {J.}~\bibnamefont
  {Zeng}}, \bibinfo {author} {\bibfnamefont {J.~W.}\ \bibnamefont {Van~Orden}},
  \ and\ \bibinfo {author} {\bibfnamefont {W.}~\bibnamefont {Roberts}},\ }\href
  {\doibase 10.1103/PhysRevD.52.5229} {\bibfield  {journal} {\bibinfo
  {journal} {Phys. Rev. D}\ }\textbf {\bibinfo {volume} {52}},\ \bibinfo
  {pages} {5229} (\bibinfo {year} {1995})},\ \Eprint
  {http://arxiv.org/abs/hep-ph/9412269} {arXiv:hep-ph/9412269} \BibitemShut
  {NoStop}%
\bibitem [{\citenamefont {Ebert}\ \emph {et~al.}(2003)\citenamefont {Ebert},
  \citenamefont {Faustov},\ and\ \citenamefont {Galkin}}]{Ebert:2002pp}%
  \BibitemOpen
  \bibfield  {author} {\bibinfo {author} {\bibfnamefont {D.}~\bibnamefont
  {Ebert}}, \bibinfo {author} {\bibfnamefont {R.~N.}\ \bibnamefont {Faustov}},
  \ and\ \bibinfo {author} {\bibfnamefont {V.~O.}\ \bibnamefont {Galkin}},\
  }\href {\doibase 10.1103/PhysRevD.67.014027} {\bibfield  {journal} {\bibinfo
  {journal} {Phys. Rev. D}\ }\textbf {\bibinfo {volume} {67}},\ \bibinfo
  {pages} {014027} (\bibinfo {year} {2003})},\ \Eprint
  {http://arxiv.org/abs/hep-ph/0210381} {arXiv:hep-ph/0210381} \BibitemShut
  {NoStop}%
\bibitem [{\citenamefont {Zouzou}\ \emph {et~al.}(1986)\citenamefont {Zouzou},
  \citenamefont {Silvestre-Brac}, \citenamefont {Gignoux},\ and\ \citenamefont
  {Richard}}]{Zouzou:1986qh}%
  \BibitemOpen
  \bibfield  {author} {\bibinfo {author} {\bibfnamefont {S.}~\bibnamefont
  {Zouzou}}, \bibinfo {author} {\bibfnamefont {B.}~\bibnamefont
  {Silvestre-Brac}}, \bibinfo {author} {\bibfnamefont {C.}~\bibnamefont
  {Gignoux}}, \ and\ \bibinfo {author} {\bibfnamefont {J.~M.}\ \bibnamefont
  {Richard}},\ }\href {\doibase 10.1007/BF01557611} {\bibfield  {journal}
  {\bibinfo  {journal} {Z. Phys. C}\ }\textbf {\bibinfo {volume} {30}},\
  \bibinfo {pages} {457} (\bibinfo {year} {1986})}\BibitemShut {NoStop}%
\bibitem [{\citenamefont {Vijande}\ \emph {et~al.}(2004)\citenamefont
  {Vijande}, \citenamefont {Fernandez}, \citenamefont {Valcarce},\ and\
  \citenamefont {Silvestre-Brac}}]{Vijande:2003ki}%
  \BibitemOpen
  \bibfield  {author} {\bibinfo {author} {\bibfnamefont {J.}~\bibnamefont
  {Vijande}}, \bibinfo {author} {\bibfnamefont {F.}~\bibnamefont {Fernandez}},
  \bibinfo {author} {\bibfnamefont {A.}~\bibnamefont {Valcarce}}, \ and\
  \bibinfo {author} {\bibfnamefont {B.}~\bibnamefont {Silvestre-Brac}},\ }\href
  {\doibase 10.1140/epja/i2003-10128-9} {\bibfield  {journal} {\bibinfo
  {journal} {Eur. Phys. J. A}\ }\textbf {\bibinfo {volume} {19}},\ \bibinfo
  {pages} {383} (\bibinfo {year} {2004})},\ \Eprint
  {http://arxiv.org/abs/hep-ph/0310007} {arXiv:hep-ph/0310007} \BibitemShut
  {NoStop}%
\bibitem [{\citenamefont {Lee}\ and\ \citenamefont {Yasui}(2009)}]{Lee:2009rt}%
  \BibitemOpen
  \bibfield  {author} {\bibinfo {author} {\bibfnamefont {S.~H.}\ \bibnamefont
  {Lee}}\ and\ \bibinfo {author} {\bibfnamefont {S.}~\bibnamefont {Yasui}},\
  }\href {\doibase 10.1140/epjc/s10052-009-1140-x} {\bibfield  {journal}
  {\bibinfo  {journal} {Eur. Phys. J. C}\ }\textbf {\bibinfo {volume} {64}},\
  \bibinfo {pages} {283} (\bibinfo {year} {2009})},\ \Eprint
  {http://arxiv.org/abs/0901.2977} {arXiv:0901.2977 [hep-ph]} \BibitemShut
  {NoStop}%
\bibitem [{\citenamefont {Feng}\ \emph {et~al.}(2013)\citenamefont {Feng},
  \citenamefont {Guo},\ and\ \citenamefont {Zou}}]{Feng:2013kea}%
  \BibitemOpen
  \bibfield  {author} {\bibinfo {author} {\bibfnamefont {G.~Q.}\ \bibnamefont
  {Feng}}, \bibinfo {author} {\bibfnamefont {X.~H.}\ \bibnamefont {Guo}}, \
  and\ \bibinfo {author} {\bibfnamefont {B.~S.}\ \bibnamefont {Zou}},\
  }\href@noop {} {\  (\bibinfo {year} {2013})},\ \Eprint
  {http://arxiv.org/abs/1309.7813} {arXiv:1309.7813 [hep-ph]} \BibitemShut
  {NoStop}%
\bibitem [{\citenamefont {Gelman}\ and\ \citenamefont
  {Nussinov}(2003)}]{Gelman:2002wf}%
  \BibitemOpen
  \bibfield  {author} {\bibinfo {author} {\bibfnamefont {B.~A.}\ \bibnamefont
  {Gelman}}\ and\ \bibinfo {author} {\bibfnamefont {S.}~\bibnamefont
  {Nussinov}},\ }\href {\doibase 10.1016/S0370-2693(02)03069-1} {\bibfield
  {journal} {\bibinfo  {journal} {Phys. Lett. B}\ }\textbf {\bibinfo {volume}
  {551}},\ \bibinfo {pages} {296} (\bibinfo {year} {2003})},\ \Eprint
  {http://arxiv.org/abs/hep-ph/0209095} {arXiv:hep-ph/0209095} \BibitemShut
  {NoStop}%
\bibitem [{\citenamefont {Navarra}\ \emph {et~al.}(2007)\citenamefont
  {Navarra}, \citenamefont {Nielsen},\ and\ \citenamefont
  {Lee}}]{Navarra:2007yw}%
  \BibitemOpen
  \bibfield  {author} {\bibinfo {author} {\bibfnamefont {F.~S.}\ \bibnamefont
  {Navarra}}, \bibinfo {author} {\bibfnamefont {M.}~\bibnamefont {Nielsen}}, \
  and\ \bibinfo {author} {\bibfnamefont {S.~H.}\ \bibnamefont {Lee}},\ }\href
  {\doibase 10.1016/j.physletb.2007.04.010} {\bibfield  {journal} {\bibinfo
  {journal} {Phys. Lett. B}\ }\textbf {\bibinfo {volume} {649}},\ \bibinfo
  {pages} {166} (\bibinfo {year} {2007})},\ \Eprint
  {http://arxiv.org/abs/hep-ph/0703071} {arXiv:hep-ph/0703071} \BibitemShut
  {NoStop}%
\bibitem [{\citenamefont {Karliner}\ and\ \citenamefont
  {Rosner}(2017)}]{Karliner:2017qjm}%
  \BibitemOpen
  \bibfield  {author} {\bibinfo {author} {\bibfnamefont {M.}~\bibnamefont
  {Karliner}}\ and\ \bibinfo {author} {\bibfnamefont {J.~L.}\ \bibnamefont
  {Rosner}},\ }\href {\doibase 10.1103/PhysRevLett.119.202001} {\bibfield
  {journal} {\bibinfo  {journal} {Phys. Rev. Lett.}\ }\textbf {\bibinfo
  {volume} {119}},\ \bibinfo {pages} {202001} (\bibinfo {year} {2017})},\
  \Eprint {http://arxiv.org/abs/1707.07666} {arXiv:1707.07666 [hep-ph]}
  \BibitemShut {NoStop}%
\bibitem [{\citenamefont {Maiani}\ \emph {et~al.}(2023)\citenamefont {Maiani},
  \citenamefont {Pilloni}, \citenamefont {Polosa},\ and\ \citenamefont
  {Riquer}}]{Maiani:2022qze}%
  \BibitemOpen
  \bibfield  {author} {\bibinfo {author} {\bibfnamefont {L.}~\bibnamefont
  {Maiani}}, \bibinfo {author} {\bibfnamefont {A.}~\bibnamefont {Pilloni}},
  \bibinfo {author} {\bibfnamefont {A.~D.}\ \bibnamefont {Polosa}}, \ and\
  \bibinfo {author} {\bibfnamefont {V.}~\bibnamefont {Riquer}},\ }\href
  {\doibase 10.1016/j.physletb.2022.137624} {\bibfield  {journal} {\bibinfo
  {journal} {Phys. Lett. B}\ }\textbf {\bibinfo {volume} {836}},\ \bibinfo
  {pages} {137624} (\bibinfo {year} {2023})},\ \Eprint
  {http://arxiv.org/abs/2208.02730} {arXiv:2208.02730 [hep-ph]} \BibitemShut
  {NoStop}%
\bibitem [{\citenamefont {Yan}\ and\ \citenamefont
  {Valderrama}(2022)}]{Yan:2021wdl}%
  \BibitemOpen
  \bibfield  {author} {\bibinfo {author} {\bibfnamefont {M.-J.}\ \bibnamefont
  {Yan}}\ and\ \bibinfo {author} {\bibfnamefont {M.~P.}\ \bibnamefont
  {Valderrama}},\ }\href {\doibase 10.1103/PhysRevD.105.014007} {\bibfield
  {journal} {\bibinfo  {journal} {Phys. Rev. D}\ }\textbf {\bibinfo {volume}
  {105}},\ \bibinfo {pages} {014007} (\bibinfo {year} {2022})},\ \Eprint
  {http://arxiv.org/abs/2108.04785} {arXiv:2108.04785 [hep-ph]} \BibitemShut
  {NoStop}%
\bibitem [{\citenamefont {Takizawa}\ and\ \citenamefont
  {Takeuchi}(2013)}]{Takizawa:2012hy}%
  \BibitemOpen
  \bibfield  {author} {\bibinfo {author} {\bibfnamefont {M.}~\bibnamefont
  {Takizawa}}\ and\ \bibinfo {author} {\bibfnamefont {S.}~\bibnamefont
  {Takeuchi}},\ }\href {\doibase 10.1093/ptep/ptt063} {\bibfield  {journal}
  {\bibinfo  {journal} {PTEP}\ }\textbf {\bibinfo {volume} {2013}},\ \bibinfo
  {pages} {093D01} (\bibinfo {year} {2013})},\ \Eprint
  {http://arxiv.org/abs/1206.4877} {arXiv:1206.4877 [hep-ph]} \BibitemShut
  {NoStop}%
\bibitem [{\citenamefont {Ferretti}\ \emph {et~al.}(2014)\citenamefont
  {Ferretti}, \citenamefont {Galat\`a},\ and\ \citenamefont
  {Santopinto}}]{Ferretti:2014xqa}%
  \BibitemOpen
  \bibfield  {author} {\bibinfo {author} {\bibfnamefont {J.}~\bibnamefont
  {Ferretti}}, \bibinfo {author} {\bibfnamefont {G.}~\bibnamefont {Galat\`a}},
  \ and\ \bibinfo {author} {\bibfnamefont {E.}~\bibnamefont {Santopinto}},\
  }\href {\doibase 10.1103/PhysRevD.90.054010} {\bibfield  {journal} {\bibinfo
  {journal} {Phys. Rev. D}\ }\textbf {\bibinfo {volume} {90}},\ \bibinfo
  {pages} {054010} (\bibinfo {year} {2014})},\ \Eprint
  {http://arxiv.org/abs/1401.4431} {arXiv:1401.4431 [nucl-th]} \BibitemShut
  {NoStop}%
\bibitem [{\citenamefont {Guo}\ \emph {et~al.}(2015)\citenamefont {Guo},
  \citenamefont {Hanhart}, \citenamefont {Kalashnikova}, \citenamefont
  {{Mei{\ss}ner}},\ and\ \citenamefont {Nefediev}}]{Guo:2014taa}%
  \BibitemOpen
  \bibfield  {author} {\bibinfo {author} {\bibfnamefont {F.-K.}\ \bibnamefont
  {Guo}}, \bibinfo {author} {\bibfnamefont {C.}~\bibnamefont {Hanhart}},
  \bibinfo {author} {\bibfnamefont {{\relax Yu}.~S.}\ \bibnamefont
  {Kalashnikova}}, \bibinfo {author} {\bibfnamefont {U.-G.}\ \bibnamefont
  {{Mei{\ss}ner}}}, \ and\ \bibinfo {author} {\bibfnamefont {A.~V.}\
  \bibnamefont {Nefediev}},\ }\href {\doibase 10.1016/j.physletb.2015.02.013}
  {\bibfield  {journal} {\bibinfo  {journal} {Phys. Lett.}\ }\textbf {\bibinfo
  {volume} {B742}},\ \bibinfo {pages} {394} (\bibinfo {year} {2015})},\ \Eprint
  {http://arxiv.org/abs/1410.6712} {arXiv:1410.6712 [hep-ph]} \BibitemShut
  {NoStop}%
\bibitem [{\citenamefont {Esposito}\ \emph {et~al.}(2022)\citenamefont
  {Esposito}, \citenamefont {Maiani}, \citenamefont {Pilloni}, \citenamefont
  {Polosa},\ and\ \citenamefont {Riquer}}]{Esposito:2021vhu}%
  \BibitemOpen
  \bibfield  {author} {\bibinfo {author} {\bibfnamefont {A.}~\bibnamefont
  {Esposito}}, \bibinfo {author} {\bibfnamefont {L.}~\bibnamefont {Maiani}},
  \bibinfo {author} {\bibfnamefont {A.}~\bibnamefont {Pilloni}}, \bibinfo
  {author} {\bibfnamefont {A.~D.}\ \bibnamefont {Polosa}}, \ and\ \bibinfo
  {author} {\bibfnamefont {V.}~\bibnamefont {Riquer}},\ }\href {\doibase
  10.1103/PhysRevD.105.L031503} {\bibfield  {journal} {\bibinfo  {journal}
  {Phys. Rev. D}\ }\textbf {\bibinfo {volume} {105}},\ \bibinfo {pages}
  {L031503} (\bibinfo {year} {2022})},\ \Eprint
  {http://arxiv.org/abs/2108.11413} {arXiv:2108.11413 [hep-ph]} \BibitemShut
  {NoStop}%
\bibitem [{\citenamefont {Dai}\ \emph {et~al.}(2022)\citenamefont {Dai},
  \citenamefont {Molina},\ and\ \citenamefont {Oset}}]{Dai:2021vgf}%
  \BibitemOpen
  \bibfield  {author} {\bibinfo {author} {\bibfnamefont {L.~R.}\ \bibnamefont
  {Dai}}, \bibinfo {author} {\bibfnamefont {R.}~\bibnamefont {Molina}}, \ and\
  \bibinfo {author} {\bibfnamefont {E.}~\bibnamefont {Oset}},\ }\href {\doibase
  10.1103/PhysRevD.105.016029} {\bibfield  {journal} {\bibinfo  {journal}
  {Phys. Rev. D}\ }\textbf {\bibinfo {volume} {105}},\ \bibinfo {pages}
  {016029} (\bibinfo {year} {2022})},\ \Eprint
  {http://arxiv.org/abs/2110.15270} {arXiv:2110.15270 [hep-ph]} \BibitemShut
  {NoStop}%
\bibitem [{\citenamefont {Padmanath}\ and\ \citenamefont
  {Prelovsek}(2022)}]{Padmanath:2022cvl}%
  \BibitemOpen
  \bibfield  {author} {\bibinfo {author} {\bibfnamefont {M.}~\bibnamefont
  {Padmanath}}\ and\ \bibinfo {author} {\bibfnamefont {S.}~\bibnamefont
  {Prelovsek}},\ }\href {\doibase 10.1103/PhysRevLett.129.032002} {\bibfield
  {journal} {\bibinfo  {journal} {Phys. Rev. Lett.}\ }\textbf {\bibinfo
  {volume} {129}},\ \bibinfo {pages} {032002} (\bibinfo {year} {2022})},\
  \Eprint {http://arxiv.org/abs/2202.10110} {arXiv:2202.10110 [hep-lat]}
  \BibitemShut {NoStop}%
\bibitem [{\citenamefont {Lyu}\ \emph {et~al.}(2023)\citenamefont {Lyu},
  \citenamefont {Aoki}, \citenamefont {Doi}, \citenamefont {Hatsuda},
  \citenamefont {Ikeda},\ and\ \citenamefont {Meng}}]{Lyu:2023xro}%
  \BibitemOpen
  \bibfield  {author} {\bibinfo {author} {\bibfnamefont {Y.}~\bibnamefont
  {Lyu}}, \bibinfo {author} {\bibfnamefont {S.}~\bibnamefont {Aoki}}, \bibinfo
  {author} {\bibfnamefont {T.}~\bibnamefont {Doi}}, \bibinfo {author}
  {\bibfnamefont {T.}~\bibnamefont {Hatsuda}}, \bibinfo {author} {\bibfnamefont
  {Y.}~\bibnamefont {Ikeda}}, \ and\ \bibinfo {author} {\bibfnamefont
  {J.}~\bibnamefont {Meng}},\ }\href@noop {} {\  (\bibinfo {year} {2023})},\
  \Eprint {http://arxiv.org/abs/2302.04505} {arXiv:2302.04505 [hep-lat]}
  \BibitemShut {NoStop}%
\bibitem [{\citenamefont {Ke}\ \emph {et~al.}(2022)\citenamefont {Ke},
  \citenamefont {Liu},\ and\ \citenamefont {Li}}]{Ke:2021rxd}%
  \BibitemOpen
  \bibfield  {author} {\bibinfo {author} {\bibfnamefont {H.-W.}\ \bibnamefont
  {Ke}}, \bibinfo {author} {\bibfnamefont {X.-H.}\ \bibnamefont {Liu}}, \ and\
  \bibinfo {author} {\bibfnamefont {X.-Q.}\ \bibnamefont {Li}},\ }\href
  {\doibase 10.1140/epjc/s10052-022-10092-8} {\bibfield  {journal} {\bibinfo
  {journal} {Eur. Phys. J. C}\ }\textbf {\bibinfo {volume} {82}},\ \bibinfo
  {pages} {144} (\bibinfo {year} {2022})},\ \Eprint
  {http://arxiv.org/abs/2112.14142} {arXiv:2112.14142 [hep-ph]} \BibitemShut
  {NoStop}%
\bibitem [{\citenamefont {Albaladejo}(2022)}]{Albaladejo:2021vln}%
  \BibitemOpen
  \bibfield  {author} {\bibinfo {author} {\bibfnamefont {M.}~\bibnamefont
  {Albaladejo}},\ }\href {\doibase 10.1016/j.physletb.2022.137052} {\bibfield
  {journal} {\bibinfo  {journal} {Phys. Lett. B}\ }\textbf {\bibinfo {volume}
  {829}},\ \bibinfo {pages} {137052} (\bibinfo {year} {2022})},\ \Eprint
  {http://arxiv.org/abs/2110.02944} {arXiv:2110.02944 [hep-ph]} \BibitemShut
  {NoStop}%
\bibitem [{\citenamefont {Du}\ \emph {et~al.}(2022)\citenamefont {Du},
  \citenamefont {Baru}, \citenamefont {Dong}, \citenamefont {Filin},
  \citenamefont {Guo}, \citenamefont {Hanhart}, \citenamefont {Nefediev},
  \citenamefont {Nieves},\ and\ \citenamefont {Wang}}]{Du:2021zzh}%
  \BibitemOpen
  \bibfield  {author} {\bibinfo {author} {\bibfnamefont {M.-L.}\ \bibnamefont
  {Du}}, \bibinfo {author} {\bibfnamefont {V.}~\bibnamefont {Baru}}, \bibinfo
  {author} {\bibfnamefont {X.-K.}\ \bibnamefont {Dong}}, \bibinfo {author}
  {\bibfnamefont {A.}~\bibnamefont {Filin}}, \bibinfo {author} {\bibfnamefont
  {F.-K.}\ \bibnamefont {Guo}}, \bibinfo {author} {\bibfnamefont
  {C.}~\bibnamefont {Hanhart}}, \bibinfo {author} {\bibfnamefont
  {A.}~\bibnamefont {Nefediev}}, \bibinfo {author} {\bibfnamefont
  {J.}~\bibnamefont {Nieves}}, \ and\ \bibinfo {author} {\bibfnamefont
  {Q.}~\bibnamefont {Wang}},\ }\href {\doibase 10.1103/PhysRevD.105.014024}
  {\bibfield  {journal} {\bibinfo  {journal} {Phys. Rev. D}\ }\textbf {\bibinfo
  {volume} {105}},\ \bibinfo {pages} {014024} (\bibinfo {year} {2022})},\
  \Eprint {http://arxiv.org/abs/2110.13765} {arXiv:2110.13765 [hep-ph]}
  \BibitemShut {NoStop}%
\bibitem [{\citenamefont {Jia}\ \emph {et~al.}(2023)\citenamefont {Jia},
  \citenamefont {Yan}, \citenamefont {Zhang}, \citenamefont {Shi},
  \citenamefont {Li},\ and\ \citenamefont {Guo}}]{Jia:2022qwr}%
  \BibitemOpen
  \bibfield  {author} {\bibinfo {author} {\bibfnamefont {Z.-S.}\ \bibnamefont
  {Jia}}, \bibinfo {author} {\bibfnamefont {M.-J.}\ \bibnamefont {Yan}},
  \bibinfo {author} {\bibfnamefont {Z.-H.}\ \bibnamefont {Zhang}}, \bibinfo
  {author} {\bibfnamefont {P.-P.}\ \bibnamefont {Shi}}, \bibinfo {author}
  {\bibfnamefont {G.}~\bibnamefont {Li}}, \ and\ \bibinfo {author}
  {\bibfnamefont {F.-K.}\ \bibnamefont {Guo}},\ }\href {\doibase
  10.1103/PhysRevD.107.074029} {\bibfield  {journal} {\bibinfo  {journal}
  {Phys. Rev. D}\ }\textbf {\bibinfo {volume} {107}},\ \bibinfo {pages}
  {074029} (\bibinfo {year} {2023})},\ \Eprint
  {http://arxiv.org/abs/2211.02479} {arXiv:2211.02479 [hep-ph]} \BibitemShut
  {NoStop}%
\end{thebibliography}

%

\end{document}